\documentclass[12pt]{article}
\usepackage{latexsym}

\newcommand{\re}{{\rm Re}}
\newcommand{\Omk}{\Omega_{k0}}
\newcommand{\Omkt}{\Omega_k(\tau)}

\newcommand{\bea}{\begin{eqnarray}}
\newcommand{\eea}{\end{eqnarray}}
\newcommand{\be}{\begin{equation}}
\newcommand{\ee}{\end{equation}}
\newcommand{\pkt}{\; .}
\newcommand{\pu}{\; .}
\newcommand{\kma}{\; ,}
\newcommand{\ko}{\; ,}

\newcommand{\calft} { \tilde{\cal  F} }
\newcommand{\calf} { {\cal  F} }
\newcommand{\caln}{{\cal N}}

\newcommand{\calC}{{\cal C}}

\newcommand{\intk}{\int\!
\frac{d^{n-1}k}{(2\pi)^{n-1}}}
\newcommand{\intkd}{\int\!
\frac{d^{3}k}{(2\pi)^{3}}}
\newcommand{\bfk}{{\bf k}}
\newcommand{\bfx}{{\bf x}}
\begin{document}
\begin{titlepage}
\begin{flushright}
DO-TH-99/09\\
June 1999
\end{flushright}
\vspace{20mm}
\begin{center}
{\Large \bf  Out-of-equilibrium evolution of scalar fields in
FRW cosmology: renormalization
and numerical simulations}\\
\vspace{10mm}
{\large  J\"urgen Baacke\footnote{
 e-mail:~baacke@physik.uni-dortmund.de} and 
Carsten P\"atzold\footnote{e-mail:~paetzold@hal1.physik.uni-dortmund.de} 
\vspace{15mm}}\\
{\large Institut f\"ur Physik, Universit\"at Dortmund} \\
{\large D - 44221 Dortmund , Germany}
\vspace{10mm}

{\bf Abstract}
\end{center}
We present a renormalized computational framework
for the evolution of a self-interacting scalar field (inflaton) 
and its quantum fluctuations in an FRW background geometry. We
include a coupling of the field to the Ricci scalar with a general
coupling parameter $\xi$.   
We take into account the classical and quantum
back reactions, i.e., we consider the the dynamical evolution of the 
cosmic scale 
 factor. We perform, in  the one-loop and in the large-$N$ 
approximation, the renormalization of the equation of motion
for the inflaton field, and of its energy momentum tensor.
Our formalism is based on a perturbative expansion for the mode functions, 
and uses dimensional regularization. The renormalization procedure
is manifestly covariant and the counter terms are independent 
of the initial state. Some shortcomings in the renormalization
of the energy-momentum tensor in an earlier publication
are corrected. We avoid the 
occurence of initial singularities 
by constructing a suitable class of initial states.  
The formalism is implemented numerically and we present some
results for the evolution in the post-inflationary preheating
era.
\end{titlepage}

\section{Introduction}
Nonequilibrium processes in cosmology have recently been considered
 by various authors. The main interest has been
centered around the possible inflationary period of the universe
(see e.g. \cite{Abbott:1986,Kolb:1990,Linde:1990})
and the subsequent reheating \cite{Dolgov:1982,Abbott:1982}.
It has already been found, by considering parametric
resonance \cite{Kofman:1994}
associated with oscillations of the inflaton field
and by exact computations including back reaction
both in Minkowski space 
\cite{Boya:1994,Cooper:1994} and in an expanding universe
\cite{Boya:1997C1,Boya:1997C2,Kofman:1997,Ramsey:1997}, that
the time-dependent inflaton field produces particles or
classical fluctuations preferentially
at low momenta and not in a distribution corresponding to
thermal equilibrium. 
 This process of particle
production has therefore been termed preheating
\cite{Kofman:1994,Shtanov:1995}.

The equations of motion for nonequilibrium systems
have been presented by various authors 
\cite{Jordan:1986,Calzetta:1987} using
the CTP formalism introduced by Schwinger \cite{Schwinger:1961} 
and Keldysh \cite{Keldysh:1964}. Their application to inflation
within a conformally flat Friedmann-Robertson-Walker (FRW) universe has been
initiated by Ringwald 
\cite{Ringwald:1987};
they have recently been implemented numerically
in \cite{Boya:1997C1,Boya:1997C2,Boya:1998I}
 and \cite{Ramsey:1997}.
Similar computations have been 
performed in configuration space
for the case that the fluctuations are
treated as classical ones 
\cite{Khlebnikov:1996}. Again fluctuations with rather
low momenta are strongly excited, thus justifying the
classical approximation.
Apart form such exact numerical computations there exist
also various analyses based on analytical approximations to the
solution of the mode equations 
\cite{Kofman:1997,Son:1996,Greene:1997a,Kofman:1996,Kaiser:1996}.

We have recently considered \cite{Baacke:1997c}
nonequilibrium dynamics of a scalar
inflaton field and its quantum fluctuations in FRW
cosmology. This work was  based on a method for
renormalized numerical computations in quantum field theory,
introduced for nonequilibrium dynamics in Minkowski space 
in \cite{Baacke:1997a}. In this method the leading, divergent
parts of the fluctuation integrals appearing in the
inflaton equation of motion and in the energy-momentum tensor
are separated from the numerical computation of finite
parts. This allows for a free choice of regularization,
which is performed analytically. In \cite{Baacke:1997c}
we chose dimensional regularization.
The renormalization then can be performed in the usual
way, with the standard counter terms of equilibrium
quantum field theory, manifestly covariant
\footnote{This is to be understood in the restricted sense of
special relativistic covariance.} and  independent of the
initial conditions.
In applying this method to a
scalar field in FRW space-time we found that renormalization
introduces singularities in the time variable at the
initial time $t_0$, taken to be $0$ in the following.
While the equation of motion for the inflaton field is finite
as $t\to 0$, the energy-momentum tensor is not. So the Friedmann
equations have an initial singularity. We, therefore, were not
able to numerically implement our formalism.

Depending on the parameters and the model under consideration,
these singularities can be more or less pronounced. It is well
known that the main excitation of fluctuations appears in a
low momentum resonance band,  so the ultraviolet divergences,
the precise handling of renormalization and thereby the initial
singularities may be considered 
relatively unimportant. Certainly the main 
features of inflaton dynamics, as 
presented in \cite{Boya:1997C2} will not be changed
if renormalization is handled in a more meticulous way.
However, in order to be on safe grounds, one should be
sure that the renormalization aspects can be handled properly and
consistently.

In \cite{Ramsey:1997} renormalization was performed by
adiabatic regularization, i.e., by subtracting the leading
adiabatic orders of the fluctuation terms. The problem of 
initial singularities there was avoided by starting the system 
with a smooth transition, for the inflaton mean field
and its fluctuations, from an initial phase in which the
back reaction of the fluctuations is switched off. In this initial phase
the time dependence of  cosmic scale parameter is   
fixed, and the evolution of the fluctuations is handled
analytically. The initial singularities are indeed  related to
a noncontinuos behaviour of the effective mass of the scalar field
fluctuations, a mass whose variation is determined by the background
metric and the inflaton mean field. In \cite{Baacke:1998a}
(see also \cite{Cooper:1987})
we found a different solution of the problem of initial singularities,
a Bogoliubov transformation of the initial state. 
Using our mode expansion and our construction of the initial state, 
it was shown in \cite{Lindig:1998} that this initial state 
corresponds, in the terminology of the 
adiabatic expansion, to a vacuum state of adiabatic order $4$. 

Having solved the problem of initial singularities we are
now able to implement our formalism numerically.
Besides these numerical computations with nonsingular
initial conditions we will present
here some further formal developments.

Part of these is made necessary by fact that in our
previous work we omitted several finite terms in the 
renormalized energy-momentum tensor, as 
criticized in \cite{Lindig:1998}. In applying 
dimensional regularization we did not take into account
the dimensional continuation of the 
conformal rescaling of the fields, and of the dimensional
continuation of the various basic tensors. Furthermore,
the determination of the counter terms was based on
the consideration of the conformally flat
FRW metric. It is not possible, then, to fix these
counter terms in an unique way. As a consequence 
the anomaly of the stress tensor did not appear at all.
This is, therefore, not due to a shortcoming of our formalism,
but to the very special nature of the problem.   

A further extension of our previous work is the
consideration of the large-$N$ limit in the
$O(N)$ $\sigma$ model. This requires a reconsideration
of the renormalization procedure, leading
to modified counter terms. 

The analysis presented here is performed for arbitrary
values of the conformal coupling $\xi$, which appears 
in various renormalization constants and is itself renormalized.

We will present the basic relations of FRW cosmology
in section 2. The nonequilibrium dynamics of a scalar 
mean field and its fluctuations in this geometry 
is introduced in section 3, the associated energy-momentum
tensor in section 4. In section 5 we describe our
perturbative expansion of the mode functions and derive
some expressions occuring in the various fluctuation integrals.
Renormalization of the one-loop equations of motion and 
of the energy-momentum tensor is considered in sections 6 and
7, respectively. In section 8 we construct
the Bogoliubov transformation by which
the initial singularities are removed. In section 9 we extend
our formalism to the $O(N)$ $\sigma$ model.
We present and discuss some results of our numerical
computations in section 10.

 
\section{FRW cosmology}
\setcounter{equation}{0}
We consider the Friedmann-Robertson-Walker metric
with curvature parameter $k=0$, i. e. a spatially isotropic and flat
space-time.
The line-element is given in this case by
\begin{equation} 
\label{line}
{d}s^2={d}t^2-a^2(t){d}{\bfx} ^2\; .
\end{equation}
The time evolution of the cosmic scale factor $a(t)$ is governed
by Einstein's field equation
\be\label{Einfield}
(1+\delta Z) G_{\mu\nu}+
\delta\alpha\,^{(1)}H\,_{\mu\nu}+\delta\beta\,^{(2)}H
\,_{\mu\nu}
+\delta\gamma\,^{}H\,_{\mu\nu}+\delta\Lambda g\,_{\mu\nu}=-\kappa\langle
T_{\mu\nu}
\rangle
\ee
with $\kappa= 8\pi G$. 

The Einstein curvature tensor $G_{\mu\nu}$ is given by
\begin{equation} 
G_{\mu\nu}=R_{\mu\nu}-\textstyle{\frac 1 2} g_{\mu\nu}R\; .
\end{equation}
The Ricci tensor and the Ricci scalar are defined as
\begin{eqnarray}
R_{\mu\nu}&=&R^{\lambda}_{\mu\nu\lambda}\; ,\\
R&=&g^{\mu\nu}R_{\mu\nu}\; ,
\end{eqnarray}
where
\begin{equation} 
R^{\lambda}_{\alpha\beta\gamma}=
\partial_{\gamma}\Gamma^{\lambda}_{\alpha\beta}
-\partial_{\alpha}\Gamma^{\lambda}_{\gamma\beta}
\Gamma^{\lambda}_{\gamma\sigma}
\Gamma^{\sigma}_{\alpha\beta}
-\Gamma^{\lambda}_{\alpha\sigma}\Gamma^{\sigma}_{\gamma\beta}\; .
\end{equation}
The tensors  $\,^{(1)}H_{\mu\nu}, \,^{(2)}H_{\mu\nu}$,
and $H_{\mu\nu}$ arise from the variation
of terms proportional to $R^2,R\,^{\alpha\beta}R_{\alpha\beta}$,
and $R\,^{\alpha\beta\gamma\delta}R_{\alpha\beta\gamma\delta}$ 
in the Hilbert-Einstein action.
Their precise definitions, valid for general dimension
$n$, and the subsequent identities for $n=4$ are given in
\cite{Birrell:1982}:
\bea
^{(1)}H\,_{\mu\nu}&&=\frac{1}{\sqrt{-g}}\frac{\delta}{\delta g\,^{\mu\nu}}
\int\!d^nx \sqrt{-g}\,R^2\nonumber\\
&&=2\,R_{;\mu\nu}-2g\,_{\mu\nu}\;\Box R
-\frac 1 2 g\,_{\mu\nu}\,R^2+2RR\,_{\mu\nu}\ko\\
^{(2)}H\,_{\mu\nu}&&=\frac{1}{\sqrt{-g}}\frac{\delta}{\delta g\,^{\mu\nu}}
\int\!{d}^nx \sqrt{-g}\,R\,^{\alpha\beta}R\,_{\alpha\beta}\nonumber\\
&&=2 R\,^{\alpha}_{\mu\, ;\nu\alpha}- 
\Box R_{\mu\nu}-\frac 1 2g\,_{\mu\nu}\,\Box R
+2R\,_{\mu}^{\alpha}R\,_{\alpha\nu}-
\frac{1}{2}g\,_{\mu\nu}\,R\,^{\alpha\beta}
R\,_{\alpha\beta}\nonumber\\
&&=R\,_{;\mu\nu}-\frac{1}{2}g\,_{\mu\nu}\,\Box R-
\Box R\,_{\mu\nu}-\frac 1 2 g\,_{\mu\nu}
\,R\,^{\alpha\beta}R\,_{\alpha\beta}
+2R\,^{\alpha\beta}R\,_{\alpha\mu\beta\nu}\kma \nonumber\\\\
H\,_{\mu\nu}&&=\frac{1}{\sqrt{-g}}\frac{\delta}{\delta g\,^{\mu\nu}}
\int\!{d}^nx \sqrt{-g}\,R\,^{\alpha\beta\gamma\delta}
R\,_{\alpha\beta\gamma\delta}\nonumber\\
&&=
-\frac 1 2 g\,_{\mu\nu}R\,^{\alpha\beta\gamma\delta}
R_{\alpha\beta\gamma\delta}+2
R\,^{\mu\alpha\beta\nu}R\,_{\nu}^{\alpha\beta\gamma}
-4\Box R\,_{\mu\nu}+2R\,_{;\mu\nu}\nonumber\\
&&
-4R\,_{\mu\alpha}R\,^{\alpha}_{\nu}+4R\,^{\alpha\beta}
R\,_{\alpha\mu\beta\nu}\pkt
\eea
In the case $n=4$ the generalized Gauss-Bonnet theorem states that
\be
\int\!d^4x \sqrt{-g}\,\left(R\,_{\alpha\beta\gamma\delta}
R\,^{\alpha\beta\gamma\delta}
+R^2-4R\,_{\alpha\beta}R^{\alpha\beta}\right)
\ee
is a topogical invariant. It then follows that
\be \label{hmunu}
H\,_{\mu\nu}=-^{(1)}H\,_{\mu\nu}+4 \;^{(2)}H\,_{\mu\nu}\mbox{ , }n=4\pu
\ee
Furthermore, in conformally flat space-time 
as considered here,
\be \label{2hmunu}
^{(2)}H\,_{\mu\nu}=\frac 1 3\,^{(1)}H\,_{\mu\nu}\mbox{ , }n=4\pu
\ee
These terms, as well as the terms $\delta Z G_{\mu\nu}$    and
$\delta \Lambda g_{\mu\nu}$ are introduced for the purpose of 
renormalization. They are going to absorb the divergences arising in
the energy-momentum tensor. It will be more convenient, later on,
to add analogous terms  to the
energy-momentum tensor, with coefficients
$\delta \tilde \Lambda= \delta \Lambda/\kappa$,
$\delta \tilde  \alpha=\delta \alpha/\kappa$ etc..
 Of course the cosmological constant and
higher curvature terms may be present in the bare action already,
if required by observation. Dividing (\ref{Einfield}) by  $1+\delta Z$
one sees that $\delta Z$ may also be considered as renormalizing
Newton's constant.

With the FRW metric (\ref{line}) the Einstein field equations reduce
to equations for the time-time component and for
the trace of $G_{\mu\nu}$, the Friedmann
equations
\begin{eqnarray}
\label{Friedtt}
(1+\delta Z) G_{tt}+ \alpha \,^{(1)}H_{tt}+\delta\beta\,^{(2)}H_{tt}
+\delta\gamma H_{tt}+\delta\Lambda &=&-\kappa T_{tt}\; ,\\
\label{Friedtr}
(1+\delta Z)G_\mu^\mu+ \alpha \,^{(1)}H^{\mu}_\mu
+\delta\beta\,^{(2)}H_\mu^\mu
+\delta\gamma H_\mu^\mu+n\delta\Lambda &=&-\kappa T^{\mu}_{\mu}\; .
\end{eqnarray}

We now compute the various terms for the 
$n$ dimensional line element (\ref{line}). 
For the Christoffel symbols for an $n$ dimensional flat  
FRW universe one finds
\bea
\Gamma\,^{t}_{ij}=a\dot{a}\,\delta_{ij}&\mbox{ ; }&
\Gamma\,^{j}_{kt}=\Gamma\,^{j}_{tk}=\frac{\dot{a}}{a}
\,\delta^j_k\pu
\eea
The nonvanishing components of the Riemann tensor are 
\bea
R\,^t_{itk}=\ddot{a}a\,\delta_{ik}&\mbox{ , }&
R\,^t_{ijt}=-\ddot{a}a\,\delta_{ij}\ko\\
R\,^l_{ttk}=\frac{\ddot{a}}{a}\,\delta^l_k&
\mbox{ , }&R\,^l_{tjt}=-\frac{\ddot a}{a}\,\delta^l_j\pu
\eea
for those of the Ricci tensor one finds 
\bea
R\,_{tt}=(n-1)\frac{\ddot{a}}{a}&\mbox{ , }&R\,_{ij}=\left[-\ddot{a}a-(n-2)
\dot{a}^2\right]\delta_{ij}\ko
\eea
this leads to the Ricci curvature scalar
\be
R=2(n-1)\frac{\ddot{a}}{a}+(n-1)(n-2)\left(\frac{\dot{a}}{a}\right)^2\pkt
\ee
Expressed in terms of Hubble's constant
\begin{equation} 
H(t)=\frac{\dot{a}(t)}{a(t)}\; 
\end{equation}
it takes the form
\be
R=(n-1)\left(2\dot{H}+nH^2\right)\pu
\ee
The time-time components and the trace of the tensors 
$\,^{(n)}H_{\mu\nu}$ are given by
\bea \label{htt1}
^{(1)}H\,_{tt}&=&-6H\dot R +\frac 1 2 R^2- 6 H^2R
\\ \nonumber &&+ (n-4)
\left(-2H\dot R-(n+1)RH^2\right)\kma\\ 
\label{htt2}
^{(2)}H\,_{tt}&=&-2H\dot R+\frac 1 6 R^2-2H^2R+(n-4)\left(-\frac 1 2 H\dot R
\right.\\ \nonumber
&&\left.-\frac{R^2}{24(n-1)}
-\frac 1 4(n+2)H^2R+\frac 1 8 (n-1)(n-2)^2H^4\right)\kma\\
\label{htt3}
H\,_{tt}&=&-2H\dot R+\frac 1 6 R^2-2H^2R\\ \nonumber
&&+(n-4)\left(
-\frac{R^2}{6(n-1)}-H^2R+\frac 1 2(n-1)(n-2)H^4
\right)\kma
\eea
\bea \label{hmumu1}
^{(1)}H\,_{\mu}^{\mu}&=&-6\ddot R-18H\dot R+(n-4)\left(
-2\ddot R-2(n+2)H\dot R-\frac 1 2 R^2\right)\kma\\
\label{hmumu2}
^{(2)}H\,_{\mu}^{\mu}&=&-2\ddot R -6 H\dot R +(n-4)
\left(-\frac 1 2 \ddot R-\frac 1 2 (n+3)H\dot R
\right.\\ \nonumber
&&\left.-\frac{nR^2}{8(n-1)}+\frac 1 4 
(n-2)^2H^2R-\frac 1 8n(n-1)(n-2)^2H^4\right)\kma\\
\label{hmumu3}
H\,_{\mu}^{\mu}&=&-2\ddot R -6H\dot R +(n-4)
\left(-2H\dot R -\frac{R^2}{2(n-1)}
\right.\nonumber\\&&\left.+(n-2)H^2R-\frac 1 2 n (n-1)(n-2)H^4\right)
\pkt
\eea


\section{Nonequilibrium equations for the scalar field}
\setcounter{equation}{0}
The Lagrangian density of a $\phi^4$-theory in curved space-time
is given by
\begin{equation} 
{\cal L}=\sqrt{-g}\left\{\textstyle{\frac 1 2} 
\partial_\mu \Phi \partial^\mu \Phi
-\textstyle{\frac 1 2} m^2 \Phi^2
-\frac \xi 2  R \Phi^2 -\frac{\lambda}{4!} \Phi^4\right\}
\; ,\end{equation}
where $R(x)$ is the curvature scalar and $\xi$ the
 bare dimensionless parameter 
describing the coupling of the bare scalar field to the gravitational
background. 
We split the field $\Phi$ into its expectation value $\phi$
and the quantum fluctuations $\psi$:
\begin{equation} 
\Phi(\bfx,t)=\phi(t)+\psi(\bfx,t)\; ,
\end{equation}
with
\begin{equation} 
\phi(t)=\langle\Phi(\bfx,t)
\rangle=\frac{{\rm Tr}\Phi\rho(t)}{{\rm Tr}\rho(t)}
\; ,
\end{equation}
where $\rho(t)$ is the density matrix of the system which satisfies the 
Liouville equation
\begin{equation} 
i\frac{d\rho(t)}{dt}=[{\cal H}(t),\rho(t)]
\; .\end{equation}
The one-loop equation of motion of a scalar field with
$\lambda \phi^4$ interaction
has been obtained in the FRW universe by Ringwald \cite{Ringwald:1987};
we follow closely his formulation. 
The equation of motion for the classical field is
\begin{equation} 
\ddot{\phi}+ (n-1) H\dot{\phi}+(m^2+\xi R)\phi+\frac{\lambda}{6}\phi^3+
\frac{\lambda}{2}
\langle\psi^2\rangle\phi=0
\; .\end{equation}
The expectation value of the quantum fluctuations 
$\langle\psi^2\rangle$ can be expressed as
\begin{equation} 
\langle\psi^2\rangle=-i G(t,\bfx;t,\bfx)
\end{equation}
in terms of the
non-equilibrium Green function $G(t,\bfx;t',\bfx')$ which 
satisfies 
\bea
\label{comtmode}
\left[\frac{{\partial}^2}{{\partial}t^2}+(n-1)H\frac{{\partial}}
{{\partial}t}+ a^{-2}(t)\vec \nabla^2+m^2+\xi R(t)\right.&&\nonumber\\
\hspace{2cm}\left.+
\frac{\lambda}{2}\phi^2(t)\right]G(t,\bfx;t',\bfx') =
\frac{i}{a^3(t)}\delta(t,\bfx;t',\bfx')\; .&&
\eea
The boundary conditions for this Green functions will be 
given below.
Due to the presence of the term $H(t)\partial/\partial t$
the differential operator on the left hand side of this
equation is  non-hermitian.
It is made hermitian by introducing conformal time 
and appropriate scale factors. Conformal time is defined as
\begin{equation} 
\tau=\int\limits_{0}^{t}\!dt'\frac{1}{a(t')} 
\; .\end{equation}
In conformal time the line-element (\ref{line}) reads
\begin{equation} 
\label{newline}
{d}s^2=C(\tau)({d}\tau^2-{d}{\bfx} ^2)
\; ,\end{equation}
where the conformal factor $C(\tau)$ is given by
\begin{equation} 
C(\tau)=a^2(\tau)
\; .\end{equation}
We further rescale the scalar field and its quantum fluctuations 
 by introducing the dimensionless `conformal' fields
\begin{eqnarray}
\varphi(\tau)&=& a^{\frac n 2 -1}(t)\phi(t) \; , \\
\tilde{\psi}(\tau,\bfx)&=& a^{\frac n 2 -1}(t) \psi(t,\bfx) \; .
\end{eqnarray}
The Green function is rescaled accordingly via
\begin{equation} 
\tilde{G}(\bfx,\tau;\bfx ',\tau')=a^{\frac n 2 -1}(t)
a^{\frac n 2 -1}(t ')G(\bfx,t;\bfx,t')\; .
\end{equation}
The equation of motion of the classical field $\varphi(\tau)$
now becomes
\be
\varphi''(\tau)+a^2(\tau)
\left[m^2+\left(\xi-\xi_n\right)
R(\tau)\right]\varphi(\tau)+
\frac{\lambda(a(\tau)\mu)^{\epsilon}}{6}\varphi^3(\tau)=0
\ee
with
\be
\xi_n=\frac{n-2}{4(n-1)}
\kma \ee
and 
where the primes denote derivatives with respect to conformal time.
The two-point-function $\tilde{G}$ now satisfies
\begin{equation} 
\left[ \frac{\partial^2}{\partial \tau^2}-\nabla^2+M^2(\tau)\right]
\tilde{G}(\bfx
,\tau;\bfx,\tau ')=-\delta(\bfx, \tau;\bfx ',\tau ')
\; .
\end{equation}
Here $M^2(\tau)$ denotes the square of the effective mass term
the fluctuation field $\tilde \psi(\tau,\bfx)$
\begin{equation} 
\label{effm}
M^2(\tau)=a^2(\tau)\left[ m^2+\left(\xi-\xi_n\right)R(\tau)\right]+
\frac{\lambda(a(\tau)\mu)^{\epsilon}}{2}
\varphi^2(\tau)
\; .
\end{equation}
For later discussion  it is useful 
to divide $M^2(\tau)$ into the usual $4$ dimensional
part and into a part proportional to $(n-4)$:
\be
M^2(\tau)=a^2\left[ m^2+\left(\xi-\frac 1 6\right)R(\tau)\right]+
\frac{\lambda(a\mu)^{\epsilon}}{2}-\frac{n-4}{12\left(n-1\right)}a^2 R
\pkt \ee
When multiplied by a divergent factor $1/(n-4)$ the last
term yields a finite contribution.

The problem of determining the Green function is now essentially
reduced to the equivalent problem in Minkowski space.
We expand the fluctuation field in terms of the mode functions
$U_k(\tau)\exp(i\bfk \bfx)$ via
\begin{equation} \label{fieldex} 
\tilde{\psi}(\tau,\bfx) =
\int\! \frac{d^{n-1}k}{(2\pi)^{n-1}} 
\left[c(\bfk) U_k(\tau)e^{i\bfk \bfx}
+c^\dagger(\bfk)U^*_k(\tau)e^{-i\bfk \bfx}\right]
\; .\end{equation}
The functions $U_k(\tau)$ satisfy the mode equation
\begin{equation} 
U_k''(\tau)+\Omega^2_k(\tau)U_k(\tau)=0\; ,
\end{equation}
with\be
\Omega_k^2(\tau)=k^2+M^2(\tau)
\ee
We further impose  the initial conditions
\begin{eqnarray}
\label{incons}
U_k(0)=1&\mbox{ ; }&U'_k(0)=-i\Omega_k(0)
\; ,
\end{eqnarray}
with
\begin{equation} 
\Omega_k(0)=\sqrt{k^2+M_0^2}
\; .
\end{equation}
In the following we will use the 
short notation $\Omega_{k0}=\Omega_k(0)$.
The nonequilibrium Green function $\tilde{G}_k(\tau,\bfx;\tau '
,{\bfx}\,')$ can by expressed in terms of the mode functions
via
\begin{eqnarray}
\tilde{G}_k(\tau,\bfx;\tau '
,\bfx')&=&\int\! \frac{d^{n-1}k}{(2\pi)^{n-1}}
\frac{i}{2\Omega_{k0}}
\Biggl\{\theta(\tau-\tau ')U_k(\tau)U_k^{*}(\tau ')
e^{i\bfk(\bfx-\bfx')} \nonumber \\
&&\hspace{2cm}+\theta(\tau ' - \tau)U_k^{*}(\tau ')U_k(\tau)
e^{-i\bfk(\bfx-\bfx')}\Biggr\}
\; .
\end{eqnarray}
The expectation value of the fluctuation fields is given,
therefore, by the fluctuation integral
\begin{equation} \label{flucdef}
\tilde {\cal F}(\tau) =\langle \tilde{\psi}^2(\tau)\rangle=
-i\tilde G (\tau,\bfx;\tau,\bfx)=
\int\! \frac{d^{n-1}k}{(2\pi)^{n-1}}
\frac{|U_k(\tau)|^2}{2\Omega_{k0}}
\; .
\end{equation}
The unrenormalized equation of motion of the inflaton field
reads
\be
\varphi''+a^2\left[m^2+\left(\xi-\xi_n\right)
R\right]\varphi+\frac{\lambda(a\mu)^{\epsilon}}{6}\varphi^3+
\frac{\lambda(a\mu)^{\epsilon}}{2}
\varphi\calf=0
\pkt \ee
The regularization of the fluctuation integral and the renormalized
form of this equation will be discussed below.

The field expansion (\ref{fieldex}), together with the
equation of motion for the mode functions and the initial conditions,
defines a Fock space in which the initial quantum state or
density matrix can be represented.
The ground state is the conformal vacuum state \cite{Birrell:1982},
it has  been chosen previously \cite{Ringwald:1987,Boya:1997C1}
as the initial state for the nonequilibrium evolution.
As mentioned in the introduction
we have found \cite{Baacke:1998a} that for such an initial state
the energy-momentum tensor becomes singular at the 
initial time. The construction of a suitable
initial state, avoiding such singularities, will be
given below.


\section{The energy-momentum tensor}
\setcounter{equation}{0}
In order to formulate Einstein's field equation we have to
discuss the energy-momentum tensor of the scalar field in
curved space time. For a classical field it reads \cite{Birrell:1982}
\begin{eqnarray}
T_{\mu\nu}&=&(1-2\xi)\phi_{;\mu}\phi_{;\nu}
+\left(2\xi-\textstyle{\frac 1 2}\right)
g_{\mu\nu}g^{\rho\sigma}\phi_{;\rho}\phi_{;\sigma}-
2\xi\phi_{;\mu\nu}\phi\nonumber\\
&&+2\xi g_{\mu\nu}\phi\Box\phi-\xi G_{\mu\nu}\phi^2
+\textstyle{\frac 1 2} m^2g_{\mu\nu}\phi^2
+\frac{\lambda}{4!}g_{\mu\nu}\phi^4\; .
\end{eqnarray}
In the conformally flat FRW metric the energy-momentum tensor
is diagonal. One obtains for its time-time component 
and its trace 
\begin{eqnarray}
T_{tt}^{cl}&=&{\frac 1 2}\dot{\phi}^2
+{\frac 1 2} m^2\phi^2+\frac{\lambda}{4!}\phi^4
-\xi G_{tt}\phi^2+2(n-1)\xi H\phi\dot{\phi}\; ,\nonumber\\
T_{\mu}^{\mu\;cl}&=&\left[1-\frac n 2 +2(n-1)\xi\right]\dot{\phi}^2
+n\left(\frac 1 2 m^2\phi^2+\frac{\lambda}{24}\phi^4\right)
-\xi G_{\mu}^{\mu}\phi^2\nonumber\\
&&
+2(n-1)\xi\left[\ddot{\phi}+(n-1)H\phi\dot{\phi}\right]\phi\; . 
\end{eqnarray}
We again introduce conformal time and the conformal rescaling of the
fields. Furthermore, we include the quantum fluctuations of the field
$\varphi$.
The classical energy density then takes the form
\footnote{We continue to consider $T_{tt}$ instead of
$T_{\tau\tau}$ for convenience.}
\bea
T_{tt}^{\rm cl}&&=\frac{1}{a^{2-\epsilon}}\Biggl\{
\frac{1}{2a^2}\varphi'^2+\frac 1 2 m^2\varphi^2+\frac{\lambda(a\mu)^{
\epsilon}}{4!a^2}\varphi^4\nonumber\\
&&+2(n-1)\left(\xi-\xi_n\right)\left(\frac{H}{a}\varphi\varphi'-\frac 1 4 
(n-2)H^2\varphi^2\right)
\Biggr\}\pkt
\eea
The fluctuation energy density is given by
\bea
T_{tt}^{\rm q}&&=\frac{1}{a^{2-\epsilon}}
\intk\Biggl\{\frac{|U_k'|^2}{2a^2}+\frac{1}{2a^2}
\Omega_k(\tau)^2|U_k|^2\\ \nonumber
&&+(n-1)\left(\xi-\xi_n\right)
\left[\frac{H}{a}\frac{d}{d\tau}|U_k|^2-
\left(\frac 1 2 (n-2)H^2+\frac{R}{2(n-1)}
\right)
|U_k|^2\right]\Biggr\}\nonumber
\pkt\eea
We obtain for the classical and fluctuation parts of the trace
\bea
T^{{\rm cl} \, \mu}_{\mu}&=&\frac{1}{ a^{2-\epsilon}}\Biggl\{
2(n-1)\left(\xi-\xi_n\right)\left[\frac{\varphi'}{a}
-\frac 1 2 (n-2)H\varphi\right]^2\nonumber\\
&&+2(n-1)\xi\frac{\varphi\varphi''}{a^2}+n\left[\frac 1 2 m^2\varphi^2+
\frac{\lambda(a\mu)^{\epsilon}}{24a^2}\varphi^4\right]
\Biggr\}\\
T^{{\rm q} \,\mu}_{\mu}&=&\frac{1}{a^{2-\epsilon}}\intk\Biggl\{
\left[\frac{n-2}{2}-2(n-1)\xi\right]\Biggl[
-\frac{|U_k'|^2}{a^2}+\frac{\Omega_k(\tau)}{a^2}|U_k|^2\nonumber\\
&&+\frac 1 2 (n-2)\frac H a \frac{d}{d\tau}|U_k|^2-\frac 1 4 (n-2)^2\left(
H^2-\frac{R}{(n-1)(n-2)}\right)|U_k|^2
\Biggr]\nonumber\\
&&+\left(m^2+\frac{\lambda(a\mu)^{\epsilon}}{2a^2}\right)|U_k|^2
\Biggr\}
\pkt\eea
Energy density and pressure are
related to the energy-momentum tensor via
\begin{eqnarray}
T_{tt} &=& {\cal E} \nonumber\; , \\
T_\mu^\mu&=& {\cal E} -(n-1) p\; .
\end{eqnarray}   
It is straightforward to show, using the equations of motion
for the classical field and for the mode functions
 (\ref{comtmode}),
that the energy is covariantly conserved:
\begin{equation} 
{\cal E}'(\tau)/a(\tau)+(n-1)H(\tau)(p(\tau)+{\cal E}(\tau))=0\; .
\end{equation}


\section{Perturbative expansion}
\setcounter{equation}{0}
In order to prepare the renormalized version of the equations given
in the previous section we introduce a suitable expansion of the
mode functions, which was used
in \cite{Baacke:1997a,Baacke:1997b,Baacke:1998c} for the inflaton field
coupled to itself, to gauge bosons, and to fermions in Minkowski-space.	
In the context of FRW cosmology it has been used in a similar way
in \cite{Davies:1979}.
Adding the term $M_0^2$
on both sides of the mode function equation it takes 
the form
\begin{equation} 
\label{udgl}
\left[ \frac{d^2}{d\tau^2}+
\Omega_{k0}^2\right]U_k(\tau)=-V(\tau)U_k(\tau)\; ,
\end{equation}
with
\begin{eqnarray}
V(\tau)&=&M^2(\tau)-M_0^2\; ,
\nonumber\\
\Omega_{k0}&=&\left[\bfk^2+M_0^2\right]^{1/2}
\end{eqnarray}
(for the definition of $M^2(\tau)$ see eq.(\ref{effm})).
 Including
 the initial conditions (\ref{incons}) 
the mode functions satisfy the equivalent integral equation
\begin{equation} 
U_k(\tau)=e^{-i\Omega_{k0} \tau}+
\int\limits^{\infty}_{0}\!{d}\tau'
\Delta_{k,{\rm ret}}(\tau-\tau')V(\tau')U_k(\tau')\;,
\end{equation}
with
\begin{equation} 
\label{fvt}
\Delta_{k,{\rm ret}}(\tau-\tau')= -\frac{1}{\Omega_{k0}}
\Theta(\tau-\tau')\sin\left(\Omega_{k0}(\tau-\tau')\right) \; .
\end{equation}
We separate $U_k(\tau)$ into the trivial part corresponding to
the case $V(\tau)=0$ and a function $h_k(\tau)$ which represents the
reaction
to the potential by making the ansatz
\begin{equation} 
\label{ansatz}
U_k(\tau)=e^{-i\Omega_{k0} \tau}(1+h_k(\tau)) \; .
\end{equation}
$h_k(\tau)$ satisfies then the 
differential equation
\begin{equation}\label{fdiffeq}
\ddot{h}_k(\tau)-2i\Omega_k^0\dot{h}_k(\tau)=-V(\tau)(1+h_k(\tau)) 
\end{equation}
with the initial conditions $h_k(0)=\dot{h}_k(0)=0$, and the
associated integral equation
\begin{equation} \label{finteq}
h_k(\tau)=\int\limits^{\tau}_{0}\!{d}\tau'\Delta_{k,{\rm ret}}
(\tau-\tau')V(\tau')(1+h_k(\tau'))e^{i\Omega_{k0} (\tau-\tau')}\;\pkt
\end{equation}
We expand now $h_k(\tau)$ with respect to orders in $V(\tau)$
by writing
\begin{eqnarray}
\label{entwicklung}
h_k(\tau)&=& h_k^{(1)}(\tau)+h_k^{(2)}(\tau)+h_k^{(3)}(\tau) +\cdots \\
 &=& h_k^{(1)}(\tau)+h_k^{{(\overline{2})}}(\tau)
\; ,\end{eqnarray}
where $h_k^{(n)}(\tau)$ is of n'th order in $V(\tau)$ and 
$h_k^{{{(\overline{n})}}}(\tau)$
is the sum over all orders beginning with the n'th one:
\begin{equation} 
h_k^{(\overline{n})}(\tau)=\sum_{l=n}^\infty h_k^{(n)}(\tau)
\; .\end{equation}
The $h_k^{(n)}$ are obtained by iterating the integral
equation (\ref{finteq}). The function $h_k^{{(\overline{1})}}(\tau)$ is
identical to the function $h_k(\tau)$ itself which is obtained
by solving (\ref{fdiffeq}). The function
$h_k^{(\overline{2})}(\tau)$ can again be obtained
by iteration via
\begin{equation} \label{f2inteq}
h_k^{(\overline{2})}(\tau)=
\int\limits^{\tau}_{0}\!{d}\tau'\Delta_{k,{\rm ret}}
(\tau-\tau')V(\tau')
h_k^{{(\overline{1})}}(\tau')e^{i\Omega_{k0} (\tau-\tau')} \;,
\end{equation}
or by using the equivalent differential equation
\begin{equation} \label{f2diffeq}
{h''}_k^{(\overline{2})} (\tau)-2i\Omega_{k0}
{h'}_k^{(\overline{2})}(\tau)=-V(\tau)h_k^{{(\overline{1})}}(\tau) \; .
\end{equation}
This iteration has the numerical aspect that it avoids computing
$h_k^{{(\overline{2})}}$ via the small difference
$h_k^{{(\overline{1})}}-h_k^{(1)}$. 

The integral equations
are used in order to derive the asymptotic behaviour as
$\Omega_{k0}\to \infty$ and to separate divergent and finite 
contributions.
We will give here the relevant leading terms for $h_k^{(1)}(\tau)$ and
$h_k^{(2)}(\tau)$. We have
\begin{equation} 
h_k^{(1)}(\tau)=\frac{i}{2 \Omega_{k0}}
\int\limits^{\tau}_{0}\!{d}\tau'
(\exp(2 i \Omega_{k0}(\tau-\tau'))-1)V(\tau')  \; .
\end{equation}
Integrating by parts we obtain
\begin{equation} \label{f1exp}
h_k^{(1)}(\tau)=
-\frac{i}{2\Omega_{k0}}\int\limits^{\tau}_{0}\!{d}\tau'
V(\tau')-\frac{1}{4\Omega_{k0}^2}V(\tau)
+\frac{1}{4\Omega_{k0}^2}\int\limits^{\tau}_{0}\!{d}\tau'
\exp(2 i \Omega_{k0}(\tau-\tau'))V'(\tau')\;,
\end{equation}
or, by another integration by parts,
\begin{eqnarray}
h_k^{(1)}(\tau)&=&
-\frac{i}{2\Omega_{k0}}\int\limits^{\tau}_{0}\!{d}\tau'
V(\tau')-\frac{1}
{4\Omega_{k0}^2}V(\tau)+\frac{i}{8\Omega_{k0}^3} V'(\tau) \\
&&-\frac{i}{8\Omega_{k0}^3}\int\limits^{\tau}_{0}\!{d}\tau'
\exp(2 i \Omega_{k0} (\tau-\tau'))V''(\tau') \; .
\end{eqnarray}
We will need often the real part of $h^{(1)}_k$ for which we find
\be  \label{realexp}
 {\rm Re}\;h^{(1)}_k(\tau) 
=-\frac{1}{4\Omega_{k0}^2}V(\tau)
+ \frac{1}{4\Omega_{k0}^2}{\cal C}(V',\tau)\; .
\ee
Here we have introduced the notation
\footnote{For the numerical computation we use the addition theorem
to split the integral into two integrals whose integrands depend
on $t'$ only; these  can be updated easily.}
\be
{\cal C}(f,\tau)=\int\limits^{\tau}_{0}\!{d}\tau'
\cos(2  \Omega_{k0}(\tau-\tau'))f(\tau') \; .
\ee
which will prove to be useful later.
For the leading behaviour of $h_k^{(2)}(\tau)$ we find
\begin{equation} \label{f2exp}
h_k^{(2)}(\tau)= -\frac{1}{4\Omega_{k0}^2}
\int\limits^{\tau}_{0}\!{d}\tau'\int\limits^{t'}_{0}\!{d}\tau''
V(\tau')V(\tau '') + O(\Omega_{k0}^{-3}) \; .
\end{equation}
In terms of this perturbative expansion we can write the mode functions
appearing in the fluctuation integrals in the equation of motion and in
the energy-momentum tensor as
\begin{equation} 
|U_k|^2=1+2 {\rm Re }\, h_k^{(\overline{1})}+|{h_k^{(\overline{1})}}|^2
\; ,\end{equation}
and
\begin{eqnarray}
|U'_k|^2&=&
\Omega_{k0}^2\left(1+2{\rm Re }\, h_k^{(\overline{1})}+|{h_k^{(\overline{1})}}|^2\right)
+|h_k^{\prime(\overline{1})}|^2\nonumber\\
&&\hspace{1cm}-i\Omega_{k0}\left(-2i{\rm Im }\,{h'_k}^{(\overline{1})}
-2i{\rm Im }\, h_k^{(\overline{1})*}{h}_k^{\prime(
\overline{1})}\right)
\; .\end{eqnarray}
As the potential is real, the leading behaviour of the sums
is
\begin{eqnarray}
\label{reskalsum}
1+2{\rm Re }\,{h}_k^{(\overline{1})}+
|h_k^{(\overline{1})}|^2&=&1-\frac{1}{2\Omega_{k0}^2}V(\tau)+
\frac{1}{4\Omega_{k0}^3}\sin(2\Omega_{k0}\tau)V'(0)+
\frac{1}{8\Omega_{k0}^4}V''(\tau)\nonumber\\
&&\hspace{-15mm}-\frac{1}{8\Omega_{k0}^4}
\cos(2\Omega_{k0}\tau)V''(0)
+\frac{3}{8\Omega_{k0}^4}V^2(\tau)+{\cal O}(\Omega_{k0}^{-5})\;,
\end{eqnarray}
and
\begin{eqnarray}
\label{imskalsum}
-2i{\rm Im }\,{h}_k^{\prime(\overline{1})}-2i{\rm Im }\, h_k^{(\overline{1})*}{h}_k^
{\prime(\overline{1})}&=&\frac{i}{\Omega_{k0}}V(\tau)
-\frac{i}{2\Omega_{k0}^2}\sin(2\Omega_{k0}\tau)V'(0)
-\frac{i}{4\Omega_{k0}^3}V''(\tau)\nonumber\\
&&+\frac{i}{4\Omega_{k0}^3}\cos(2\Omega_{k0}\tau)V''(0)
-\frac{3i}{4\Omega_{k0}^3}V^2(\tau)+{\cal
O}(\Omega_{k0}^{-4})\; .\nonumber\\
\end{eqnarray}
From the Wronskian relation
\begin{equation} 
U_k{U_k^*}'-U_k'U_k^*=2i\Omega_{k0}
\end{equation}
we obtain 
\begin{equation} 
2i\Omega_{k0}\left(2{\rm Re }\,{h}_k^{(\overline{1})}+
|h_k^{(\overline{1})}|^2\right)
-2i{\rm Im }\,{h}_k^{\prime
(\overline{1})}-2i{\rm Im }\, h_k^{(\overline{1})*}{h}_k^{\prime(
\overline{1})}=0
\; ,
\end{equation}
which proves to be useful in simplifying the mode integrals
occuring in the energy-momentum tensor.
Using the Wronskian relation we obtain for $|U'_k|^2$
\be
\label{upsq}
|U'_k|^2=\Omega_0^2-\Omega_0^2\left(2{\rm Re }\,{h}_k^{(\overline{1})}+
|h_k^{(\overline{1})}|^2\right)+|h_k'|^2
\pkt \ee
By means of the equation of motion and Eq. (\ref{upsq}) we have
\bea
\frac 1 2 \frac{d^2}{d\tau^2}|U_k|^2&=&|U_k'|^2-\Omega(t)^2|U_k|^2
\\ \nonumber
&=&-\left(V(\tau)+2\Omega_0^2\right)\left( 2{\rm Re }\,{h}_k^{(\overline{1})}+
|h_k^{(\overline{1})}|^2\right)+|h_k'|^2-V(\tau)
\eea


\section{Renormalization of the equation of motion}
\setcounter{equation}{0}
Having expanded the mode functions perturbatively, we
are able to separate the divergent parts of the mode sum
in an analytic way, leaving the finite parts for numerical
computation. This allows for a free choice of regularizations.
Furthermore, the analytic expressions for the divergent parts,
as separated from the mode sum, essentially have the standard 
 form as obtained from Feynman graphs, so a comparison with
purely analytic approaches is straightforward.

Among the regularizations used in field theory in curved
space are: point-splitting, dimensional, and adiabatic
regularization. Point-splitting regularization is technically 
involved as it requires performing the delicate and noncovariant
limit $x' \to x$. Adiabatic regularization actually is
a subtraction, it has often been used  
\cite{Birrell:1982}, most recently in \cite{Ramsey:1997}. 
Adiabatic regularization is considered to be well suited for numerical
computations, as the entire divergent part is subtracted, encompassing
this way the problem of regularization.

Here we choose dimensional regularization as it fits in the 
most appropriate way into our formalism. We have already used
it in our previous work
on FRW cosmology. Here we have to correct for some omissions
in the finite terms, and in particular for
the conformal anomaly. 
In \cite{Baacke:1997c} we have performed
the renormalization by considering the equation of
motion for the condensate and of the Einstein
equations.
This is possible in 4-dimensional conformally flat space,
and leads to a conserved energy-momentum tensor;
it does not correspond, however, to a properly renormalized 
local action \cite{Birrell:1982} and does not 
work for more general metrics. This shortcoming has been criticized in 
\cite{Lindig:1998}.

In our previous publication we have performed the
conformal scaling in four dimensions, and applied dimensional
regularization to the resulting `flat space' equations in analogy 
to the Minkowski case. We thereby missed terms arising
from the dimensional continuation of the
conformal rescaling, resulting in particular in  the absence of some
terms involving $\log a(t)$. As to the equation of motion
for the inflaton field, these are the only corrections.
The renormalization of the energy-momentum tensor is more 
subtle and will be discussed in the next section.  

Using (\ref{realexp}) we split
the fluctuation integral of the equation of motion into a 
divergent and a convergent part.
Using (\ref{realexp}) we 
\begin{eqnarray}
\cal F(\tau)&=&\intk
\frac{1+2{\rm Re}\,h_k^{(\overline{1})}(\tau)+
|{h_k^{(\overline{1})}}(\tau)|^2}{2\Omega_{k0}}\nonumber\\
&=&\intk
\frac{1}{2\Omega_{k0}}\left(1-\frac{1}{2{\Omega_{k0}}^2}
V(\tau)+\frac{1}{2{\Omega_{k0}}^2}{\cal C}(V',\tau)\right.\nonumber\\
&&\left.\hspace{2.7cm}
+2{\rm Re }\,{h}_k^{(\overline{2})}(\tau)
+|h_k^{(\overline{1})}(\tau)|^2\right)
\; .
\end{eqnarray}
The first two terms in the integrand  have to be regularized, as they lead
to a logarithmic and a quadratic divergence. 
We can do another integration by parts of the ${\cal C}(V',\tau)$
and arrive at
\bea 
\cal F(\tau)&=&\intk
\frac{1}{2\Omega_{k0}}\Biggl\{(1-\frac{1}{2{\Omega_{k0}}^2}
V(\tau)+
\frac{1}{4\Omega_{k0}^3}\sin(2\Omega_{k0}\tau)V'(0)
\\ \nonumber
&&-\frac{1}{8\Omega_{k0}^4}
\cos(2\Omega_{k0}\tau)V''(0)
+\frac{\ddot{V}(\tau)}{8\Omega_{k0}^4}
-\frac{{\cal C}(V''',\tau)}{8\Omega_{k0}^4}
+2\re h^{{\overline(2)}}_k+|h_k|^2\Biggr\}
\pkt\eea
Those parts of the integral
that involve $\sin(2\Omega_{k0}\tau)$ and $\cos(2\Omega_{k0}\tau)$
develop a non-analytic behaviour as $\tau \to 0$. This is due to the
fact that the modulation of the integrand by the trigonometric
functions disappears in this limit. Here the non-analyticity is
of the form $\tau \log \tau$ and $\tau^2 \log \tau$, respectively.
In the energy-momentum tensor the analogous terms
result in $1/\tau$ and $\log \tau$ singularities.
These singularities will be discussed later, in section \ref{initial}.

We first rewrite
the basic equation of motion, including appropriate counter terms, as
\bea
\varphi ''+ a^2\left[m^2+\delta m+\left(\xi-\xi_n +
\delta \xi\right) R\right]\varphi \\  \nonumber
+\frac {\lambda+\delta \lambda}{6}
(a\mu)^\epsilon \varphi^3 + \frac{\lambda}{2}\varphi
(a\mu)^\epsilon {\cal F}=0\;.
\eea
Next we separate from the term $\lambda \varphi {\cal F}/2$  the
dimensionally regularized divergent parts. The two relevant expressions
behave as
\be
\label{dimreg1}
(a\mu)^\epsilon\frac{\lambda}{2}\varphi\int
\!\frac{d^{3-\epsilon}k}{(2\pi)^{3-\epsilon}}\,
\frac{1}{2\left[\bfk^2+M_0^2\right]^{1/2}}
\simeq  -  \frac{\lambda M_0^2 
\varphi}{32\pi^2}\left(L_0+1\right)
\ee
and
\be
\label{dimreg2}
-\frac{1}{8}(a\mu)
^\epsilon\lambda\varphi V(\tau)\int\!
\frac{d^{3-\epsilon} k}{(2\pi)^{3-\epsilon}}\,
\frac{1}{\left[\bfk^2+M_0^2\right]^{3/2}}
\simeq -\frac{\lambda \varphi V(\tau)}
{32\pi^2}L_0\; 
\ee
in the limit $\epsilon \to 0$.
Here we have defined
\be
L_0=\frac{2}{\epsilon}
+\ln{\frac{4\pi\mu^2a^2(\tau)}{M_0^2}}-\gamma
\pkt\ee
This cannot be absorbed, e.g., in an $\overline{MS}$ scheme,
by a renormalization counter term, as
it depends on $\tau$, and on the initial mass $M_0$.
Recalling that $V(\tau)=M^2(\tau)-M_0^2$ and 
\be
M^2(\tau)=a^2\left[ m^2+\left(\xi-\frac 1 6\right)R(\tau)\right]+
\frac{\lambda(a\mu)^{\epsilon}}{2}-\frac{n-4}{12\left(n-1\right)}a^2 R
\kma \ee
we find that the two divergent terms
combine into
\be
 -  \frac{\lambda M^2(\tau)\varphi}{32\pi^2} L_0  -  
\frac{\lambda M_0^2\varphi}{32\pi^2}
\pkt
\ee
The divergences can now be cancelled by the counter terms 
\begin{eqnarray}                 \label{deltam}
\delta m^2 &=&
\frac{\lambda m^2}{32\pi^2}L\kma \\
\label{deltalam}
\delta \lambda &=&
\frac{3\lambda^2}{32\pi^2}L\kma \\
\delta \xi& =&
\frac{\lambda(\xi-\frac 1 6)} {32\pi^2} L
\kma
\end{eqnarray}
where
\be
L=\frac{2}{\epsilon}+
\ln{\frac{4\pi\mu^2}{m^2}}-\gamma
\pkt\ee
This leaves some finite parts
proportional to $L-L_0=\ln\frac{m^2a^2}{M_0^2}$, 
and an additional finite
term due to $(n-4)$ part of $M^2(\tau)$ multiplied with the $1/(n-4)$ term
of the divergent integral.
Finally the renormalized equation of motion in $n=4$ dimensions  reads
\be
\varphi''+a^2\left[m^2+\Delta m^2+\left(\xi+\Delta\xi-\frac 1 6\right)
R\right]\varphi+\frac{\lambda+\Delta\lambda}{6}\varphi^3
+
\frac{\lambda}{2}
\varphi\hat{\calf}_{\rm fin}=0
\ee
where $\hat{\calf}_{\rm fin}$ is defined as
\begin{equation}\label{flucintfin}
\hat{{\cal F}}_{\rm fin}=-\frac{a^2R}{288\pi^2}+\calf_{\rm fin}
\ee
with
\be
\calf_{\rm fin}=-\frac{M_0^2}{16\pi^2}
+\int\! \frac{d^3k}{(2\pi)^3}\frac{1}{2\Omega_{k0}}
\left(
\frac{1}{2{\Omega_{k0}}^2}{\cal C}(V',\tau)
+2{\rm Re }\,{h}_k^{(\overline{2})}+|h_k^{(\overline{1})}|^2\right)
\pkt
\end{equation}
The finite mass and coupling constant corrections are
defined as
\begin{eqnarray}
\Delta m^2 &=&-\frac{\lambda m^2}{32\pi^2}\ln\frac{m^2a^2}{M_0^2}\; ,
\\
\Delta \lambda&=&-\frac{3\lambda^2}{32\pi^2}\ln\frac{m^2a^2}{M_0^2}\; ,
\\
\Delta \xi& =&-\frac{\lambda(\xi- \frac{1}{6})}
{32\pi^2}\ln\frac{m^2a^2}{M_0^2} \; 
\pkt \eea
Note that these are {\em time-dependent}, due to the occurence
of $a(\tau)$. This is a feature we have missed in \cite{Baacke:1997c},
as we did not continue the conformal rescaling to $n\neq 4$.


\section{Renormalization of the energy-momentum tensor}
\setcounter{equation}{0}
In order to derive a renormalized form of the Friedmann equations
we have to renormalize the energy-momentum tensor. 
In principle this 
has been discussed long ago and indeed the divergent parts 
will be found to be  in one-to-one correspondence to those
given in the standard literature \cite{Birrell:1982}. However, we have
to discuss this subject in the framework of
nonequilibrium quantum field theory,
and we are interested in particular in the precise form of the 
finite parts which will be the subject
of a numerical computation.

Our previous discussion of the renormalization of the
energy-momentum tensor had three shortcomings:

(i) as for the equation of motion we have not extended
the conformal rescaling to $n \neq 4$, these will result
in $\log a(t)$ contributions, as above ;

(ii) we have not continued the tensor structures to $n \neq 4$;
these terms, proportional to $(n-4)$,
 will be  explicitly displayed below, they contribute
finite terms when multiplied by $1/(n-4)$;

(iii) finally, we have discussed renormalization using the
Einstein equations, and not the Hilbert-Einstein action; this
is the reason why the conformal anomaly did not appear.

As, nevertheless, the energy-momentum tensor was found to 
be conserved, 
these shortcomings passed unnoticed. This fact, and the 
absence of the anomaly, are related to the very special
nature of the conformally flat metric in $n=4$.
Indeed a unique determination of the counter terms would require
the consideration of quantum field theory in a {\em general}
background metric, and will necessarily be incomplete if only
one (and even very particular) metric  is used. We will first
derive the divergent terms and discuss this rather subtle matter
afterwards.  

In deriving the divergent and finite parts of the 
energy-momentum tensor we have to consider an expansion
in powers of $(n-4)=-\epsilon$. The general expression contains 
various factors with an explicit $n$-dependence. We define
a tensor $T_{\mu\nu}^{(4)}$ where in all of these
factors $n$ is set equal to $4$, in which, however, the fluctuation
integral is defined in $n$ dimensions. 
We define a second tensor, $T^{(n-4)}_{\mu\nu}$, 
which takes into account terms
linear in $(n-4)$ from the explicit factors,
multiplied by terms proportional
to $1/(n-4)$ from the divergent fluctuation integral, and of the
counter terms. The remaining terms, less singular or with
higher powers of $(n-4)$, do not contribute when $n\to 4$. We
expand the tensor as
\be \label{tsep}
T_{\mu\nu}=T_{\mu\nu}\,^{(4)}+T_{\mu\nu}\,^{ (n-4)}
+ O(n-4)
\pkt \ee
Explicitly these tensors are given by
\begin{eqnarray}
T_{tt}\,^{(4)}&=&\frac{1}{2a^4}\varphi'^2+
\frac{1}{2a^2} \left(m^2+\delta m^2\right)\varphi^2
+\frac{\lambda+\delta \lambda}{4!a^4}\varphi^4\nonumber\\
&&+(1-6\xi-6\delta\xi)\left(
\frac{H^2}{2a^2}\varphi^2-\frac{H}{a^3}\varphi\varphi'
\right)\nonumber\\
&&+\delta \tilde{\Lambda}+\delta \tilde{\alpha} 
\,^{(1)}H^{(4)}_{tt}
+\delta \tilde{\beta} 
\,^{(2)}H^{(4)}_{tt}
+\delta \tilde{\gamma} H^{(4)}_{tt}
+\delta\tilde{Z}G^{(4)}_{tt}\nonumber
\label{engy}\\
&&+\frac{1}{a^{n-2}}\int\! \frac{d^{n-1}k}
{(2\pi)^{n-1}}\frac{1}{2\Omega_{k0}}\Biggl\{
\frac{1}{2a^2}|U_k'|^2+\frac{1}{2a^2}\Omega(\tau)|U_k|^2
-{{\frac{1}{2}}}\left(\xi -{ \frac{1}{6}}\right)R|U_k|^2\nonumber\\
&&
-{ {\frac{1}{2}} }(6\xi-1) H^2|U_k|^2+{ {\frac{1}{2}} }
(6\xi -1)\frac{H}{a}\frac{d}{d\tau}|U_k|^2
\Biggr\}\; ,\nonumber
\end{eqnarray}
\bea
T_{tt}\,^{(n-4)}&=&(n-4)
\Biggl\{6\delta\xi\frac{H^2}{4a^2}\varphi^2
\nonumber\\&&
+2\delta\xi\left(\frac{H}{a^3}\varphi\varphi'
-\frac{H^2}{2a^2}\varphi^2\right)
+\delta Z \left(-\frac 5 2 H^2\right)\nonumber\\&&
+\delta\alpha\left(-2\frac{HR'}{a}-5H^2R\right)
+\delta\gamma\left(-\frac{R^2}{18}-H^2R+3H^4\right)
\nonumber\\&&
+\delta\beta\left(\frac{HR'}{2a}-\frac{R^2}{72}-\frac{3}{2}H^2R+\frac{3}{2}H^4 \right)
\Biggr\}
\nonumber\\&&
+(n-4)\intk\frac{1}{2\Omega_0}\frac{1}{a^{2-\epsilon}}\Biggl\{
\left(2\xi-\frac 1 2 \right)\Biggl[
-\left(H^2+\frac R 6\right)|U_k|^2
\nonumber\\&&+\frac{H}{2a}\frac{d}{d\tau}|U_k|^2\Biggr]
-\left(6\xi-1\right)\left(\frac{H^2}{4}-\frac{R}{36}\right)|U_k|^2
\Biggr\}
\pkt \eea
The divergent parts of the fluctuation integral are 
\begin{eqnarray} \label{efluc}
{\cal E}_{\rm div,fluc}&=&
\frac{1}{a^{n-2}}\intk \left[\frac{\Omega_{k0}}{2a^2}
+\frac{1}{4\Omega_{k0}a^2}V(\tau)-\frac{1}{16\Omega_{k0}^3a^2}V^2(\tau)
\right.\nonumber\\
&&\hspace{2.3cm}-\frac{1}{2}(6\xi-1)\left(\frac R 6 +H^2\right)
\left[\frac{1}{2\Omega_{k0}}-
\frac{1}{4\Omega_{k0}^3}V(\tau)\right]\nonumber\\
&&\hspace{2.3cm}\left.-\frac{1}{2}(6\xi-1)\frac{H}{a}\frac{1}
{4\Omega_{k0}^3}V'(\tau)\right]\; .
\end{eqnarray}
Dimensional regularisation of the first three terms
in the integral yields
\begin{eqnarray} \label{fdiv}
&&\intk\frac{1}{a^n}\left[\frac{\Omega_{k0}}{2}
+\frac{1}{4\Omega_{k0}}V(\tau)-
\frac{1}{16\Omega_{k0}^3}V^2(\tau)\right]
\nonumber\\
&&=-\frac{M^4(\tau)}{64\pi^2a^4}L_0
+\frac{M^4(0)}{128\pi^2a^4}-\frac{M_0^2M^2(\tau)}{32\pi^2a^4}
\\
&&=-\frac{\left[m^2+\left(\xi-\xi_n\right)R 
+\frac{\lambda}{2}\varphi^2\right]^2}{64\pi^2}L_0
+\frac{M^4(0)}{128\pi^2a^4}-\frac{M_0^2M^2(\tau)}{32\pi^2a^4}\nonumber
\; .
\end{eqnarray}
The terms proportional $\lambda m^2 \varphi^2$ and
$\lambda^2/4 \varphi^4$ in (\ref{fdiv}) are cancelled by the mass
and coupling constant counter terms. The divergent term which depends on
$m^4$ determines the cosmological constant
counter term, that is
\begin{equation} 
\delta \tilde{\Lambda} =\frac{m^4}{64\pi^2} L\; .
\end{equation}
The remaining terms in (\ref{fdiv}) combine with
the corresponding expressions
of the remaining parts
of ${\cal E}_{\rm div,fluc}$:
\begin{eqnarray}\label{sdiv}
&&-\frac{1}{2}(6\xi-1)
(\frac R 6 +H^2)\frac{1}{a^{n-2}}\intk\left[\frac{1}{2\Omega_{k0}}
-\frac{1}{4\Omega_{k0}^3}V(\tau)
\right]\nonumber\\
&&=(6\xi-1)\left(\frac R 6 +H^2\right)
\frac{M^2(\tau)}{32\pi^2}L_0\nonumber\\
&&\hspace{1cm}+\frac{1}{32\pi^2}
(6\xi-1)\left(\frac R 6+H^2\right)M_0^2\;,
\end{eqnarray}
and
\bea\label{tdiv}
&&-\frac{1}{2}(6\xi-1)\frac{H}{a^{3-n}}\intk
\frac{1}{4\Omega_{k0}^3}V'(\tau) \nonumber \\
&&=
-\frac{1}{32 \pi^2}(6 \xi -1)\frac{H}{a}L_0
\Biggl[2 a H M^2+\left(\xi-\frac{1}{6}\right)a^2R'
+\lambda \varphi\varphi' -\lambda a H \varphi^2
  \nonumber \\
&&-\frac{\lambda}{2}(n-4)a H \varphi^2
-\frac{n-4}{36}\left(a^2 R' + 2 a^3 H\right)\Biggr]
+ O(n-4)
\eea
The term $\lambda(\xi-1/6)R\varphi^2$ in (\ref{fdiv})
 is cancelled by the
same term with opposite sign in (\ref{sdiv}).
The $H^2\varphi^2$-terms in (\ref{sdiv}) and (\ref{tdiv})
are absorbed by the counter term proportional to
$\delta \xi$, and the divergence
proportional to
$\varphi\varphi'$ is compensated by this counter term as well.
The remaining $\varphi$-independent but still time-dependent divergent
terms
can absorbed into the counter terms 
$\delta\tilde{\alpha} ~H_{tt}$ 
and $\delta\tilde{ Z}~ G_{tt}$.
We choose
\begin{eqnarray}
\delta \tilde{\alpha} &=&
-\frac{(\xi- \frac{1}{6})^2}{32\pi^2}
L\; , \\
\delta\tilde {Z}& =& - \frac{(\xi- \frac{1}{6} ) m^2}{16\pi^2}
L
\; .
\end{eqnarray}
This way all divergent integrals appearing in the unrenormalized
fluctuation integral of the energy are removed by the corresponding
counter terms, the renormalized expression for the energy will be 
given below.

We shall now comment on the terms leading to the trace anomaly.
Einstein's equation and therefore the energy-momentum tensor contain
the terms
\be
\delta\tilde{\alpha}\,^{(1)}H\,_{\mu\nu}
+\delta\tilde{\beta} \,^{(2)}H\,_{\mu\nu}
+\delta\tilde{\gamma}\,H\,_{\mu\nu}
\pkt \ee
All the three tensors are conserved; furthermore, in
4-dimensional conformally flat space they are linearly related
by Eqs. (\ref{hmunu}) and (\ref{2hmunu}), and are in fact
proportional to each other. Therefore, the condition of a
finite energy-momentum  tensor cannot determine the three coefficients
$\delta\tilde{\alpha}\;,\delta\tilde{\beta}$ and $\delta\tilde{\gamma}$,
if one considers just this restricted class of metrics.
So we have to reconsider our previous choice of $\delta\tilde{\alpha}$.
As we do not consider a more general metric here, we have to 
supply some additional information.

The divergent part of the energy-momentum tensor for a general
metric in dimensional regularization has been worked out
in \cite{Bunch:1979} and is discussed in detail in \cite{Birrell:1982}.
One finds
\be
\langle T_{\mu\nu}\rangle_{\rm div} =
\frac{1}{32\pi^2} \left[\frac{1}{90}H\,_{\mu\nu}
-\frac{1}{90} \,^{(2)}H\,_{\mu\nu}+
\left(\xi-\frac 1 6 \right)^2 \,^{(1)}H\,_{\mu\nu}\right]L
\pkt
\ee
As $\,^{(2)}H\,_{\mu\nu}=H\,_{\mu\nu}$, in $n=4$ and for our metric,
our divergent part is consistent with this expression. It is apparent
that within our framework there is no way 
to determine the coefficients of these
two tensors, which are the source of the conformal anomaly.
Therefore, the absence of these terms in our previous
publication is not, as suggested in \cite{Lindig:1998},
 due to an inconsistency of  our
perturbative expansion. 
They have to be taken over from a more general analysis. 
Having chosen, accordingly,
\be
\delta\tilde{\beta}=-\delta\tilde{\gamma}=\frac{1}{2880\pi^2}L
\ee
we continue from $n\neq 4$ to obtain 
the finite terms arising from an $1/(n-4)$ in 
the renormalization constants multiplied with the $(n-4)$ parts
of the tensors, as displayed explicitly in Eqs. (\ref{htt1}-\ref{hmumu3}). 
These finite parts then contain the conformal anomaly, and 
further terms proportional to $\xi -1/6$.

The anomalous parts of the zero-zero component and of the trace are
\bea
T_{tt}\,^{\rm ano}&=&\delta\tilde{\beta} \,^{(2)}H\,_{tt}
+\delta\tilde{\gamma}\,H\,_{tt}
\nonumber\\&=&\frac{1}{2880\pi^2}\left(H\frac{R'}{a}
+RH^2-\frac{1}{12}R^2-3H^4\right)
\eea
and
\bea
T_{\mu}^\mu\,^{\rm ano}&=&\delta\tilde{\beta} \,^{(2)}H\,_{\mu}^{\mu}+\delta\tilde{\gamma}
\,H\,_{\mu}^{\mu}
\nonumber\\&=&\frac{1}{2880\pi^2}\left(
\frac{R''}{a^2}+2H\frac{R'}{a}-2RH^2+12H^4
\right)
\pkt\eea

After regularization and renormalization 
the energy-momentum tensor is given by
\bea
T_{tt}^{\rm ren}&=&
\frac{1}{2a^4}\varphi'^2+\frac{1}{2a^2} (m^2+\Delta m^2)\varphi^2
+\frac{\lambda+\Delta \lambda}{4!a^4}\varphi^4
\nonumber \\
&&-6\left(\xi-{ \frac{1}{6}}+\Delta \xi\right)\left(
\frac{H^2}{2a^2}\varphi^2-\frac{H}{a^3}
\varphi\varphi'
\right)
\nonumber\\
&&+\Delta \tilde{\Lambda}+\Delta \tilde{\alpha}\,^{(1)} H_{tt}
+\Delta\tilde{Z} G_{tt} + T_{tt}^{\rm q,fin}\nonumber\\
\pkt
\eea
The renormalized expressions the tensors $\,^{(1)} H_{tt}$ 
and  $G_{tt}$ are the ones in four dimensions. 
The finite remnants of the divergent parts have the coefficients
\bea
\Delta \tilde{\alpha} &=&
\frac{\left(\xi-{ \frac{1}{6}}\right)^2}{32\pi^2}\ln{\frac{m^2a^2}{M_0^2}}\; ,
\\
\Delta \tilde{\Lambda} &=&-\frac{m^4}{64\pi^2}\ln{\frac{m^2a^2}{M_0^2}}\; ,
\\
\Delta \tilde{Z}& =&
\frac{\left(\xi-{ \frac{1}{6}}\right)m^2}{16\pi^2}\ln{\frac{m^2a^2}{M_0^2}}
\kma \end{eqnarray} 
which again are time-dependent, due to the occurence of $a(\tau)$. 

The fluctuation part of the energy density expressed through finite mode
integrals is given by
\bea \label{tttfin}
T_{tt}^{\rm q,fin}&=&{\cal E}_{\rm kin,fin}
+V(\tau)\frac{{ \cal F}_{\rm fin}}{2a^4}
+\frac{1}{2}(6\xi-1)\frac{H}{a^3} {\cal F}'_{\rm fin}
\nonumber\\
&&-\frac{1}{2}(6\xi-1)\left(\frac R 6+H^2\right)
\frac{{\cal F}_{\rm fin}}{a^2}
+T_{tt}^{\rm add}\nonumber\\
\eea
where 
\be
{\cal E}_{\rm kin,fin}(\tau)=-\frac{3M_0^4}{128\pi^2a^4}+\frac{1}{2a^4}
\intkd\frac{1}{2\Omega_0}\left[|{h_k^{(\overline{1})}}'|^2
-\frac{V^2(\tau)}{4\Omega_{k0}^2}\right]
\kma \ee
and
\bea
T_{tt}^{\rm add}&=&T_{tt}^{\rm ano}
+\frac{H}{96\pi^2a^3}\left(aHM^2-V'\right)\nonumber\\
&&+\frac{1}{16\pi^2}\left(\xi-\frac 1 6\right)\left(
\frac{\,^{(1)}H_{tt}\,^{(4)}}{36}+\frac{R^2}{72}+\frac{6H^2M^2}{a^2}\right)
\kma \eea
or, explicitly,
\bea
T_{tt}^{\rm add}&=&T_{tt}^{\rm ano}
+\frac{\left(\xi-\frac 1 6 \right)}{8\pi^2}\Biggl[-\frac{R'H}{6a}-3\left(
\xi-\frac 1 6 \right)RH^2-\frac 3 2 \frac{H^2}{a^2}\lambda\varphi^2+\frac{1}{72}R^2
\nonumber\\
&&-3m^2H^2-\frac 1 6 RH^2
\Biggr]
-\frac{\lambda}{96\pi^2a^3}H\varphi\left(\varphi'-\frac 1 2 aH\varphi\right)
-\frac{m^2}{96\pi^2}H^2\nonumber\\
\pkt\eea
Next we have to consider the renormalization of the
trace of the energy-momentum tensor.
We introduce the available counter terms into the unrenormalized
expression for
$T_{\mu}^\mu$ and separate the trace according to Eq. (\ref{tsep}).
$T_{\mu}^\mu\,^{(4)}$ is given by
\begin{eqnarray}
T_{\mu}^\mu\,^{(4)}&=&
-\left[1-6(\xi+\delta \xi)\right]\left(
\frac{\varphi'^2}{a^4}+ \frac{H^2}{a^2}\varphi^2
-2\frac{H}{a^3}\varphi\varphi'
\right)+\frac{6(\xi+\delta\xi)}{a^4}\varphi\varphi''\nonumber\\
&&+2(m^2+\delta m^2)
\frac{\varphi^2}{a^2}+\frac{\lambda+\delta \lambda}{6a^4}
\varphi^4\nonumber\\
&&+4\delta \tilde{\Lambda}+ \delta\tilde{Z} G_\mu^\mu\,^{(4)}+
\delta \tilde{\alpha} \,^{(1)} H^{\mu}_\mu\,^{(4)}+ T_{\mu}^\mu\,^{\rm ano}
\nonumber\\
&&+\frac{1}{a^{n-2}}\intk\frac{1}{2\Omega_{k0}}\left\{
(6\xi-1)\left[\left(\frac{|U_k'|^2}{a^2}
-\Omega^2(\tau)\frac{|U_k|^2}{a^2}
\right)\right.\right.\nonumber\\
&& \left.\hspace{4cm}
+\left(H^2-R/6\right)|U_k|^2-\frac{H}{a}\frac{d}{d\tau}|U_k|^2
\right]\nonumber\\
&&\left.\hspace{3cm}+\left(m^2+
\frac{\lambda}{2}\frac{\varphi^2}{a^2}\right)|U_k|^2\right\}\; .
\end{eqnarray}
We can split the trace of
the stress tensor into a divergent and into a convergent
part.
The divergent part of $T_{\mu}^\mu\,^{(4)}$ reads after dimensional
regularization
\bea
{T_{\mu}^{\mu}}_{div}&=&\frac{1}{32\pi^2a^2}(1-6\xi)\left[\frac{V''(\tau)}{a^2}
-2\frac{H}{a}V'(\tau)\right]L_0 \\ \nonumber
&&-\frac{1}{16\pi^2a^2}\left[m^2+\frac{\lambda}{2a^2}\varphi^2-(1-6\xi)
(H^2- \frac{R}{6})\right]M^2(\tau)L_0
\pkt \eea
Inserting the derivatives of the potential $V(\tau)$ 
we see that again the counter terms absorb all divergent terms in the fluctuation
integral of ${T_{\mu}^{\mu}}$.
The renormalized trace of the stress tensor takes the final form
\bea
T_{\mu}^{\mu}\,^{\rm ren}&=&
-\left[1-6(\xi+\Delta \xi)\right]\left(
\frac{\varphi'^2}{a^4}+ 
\frac{H^2}{a^2}\varphi^2-2\frac{H}{a^3}\varphi\varphi'
\right)\nonumber\\&&
+\frac{6(\xi+\Delta \xi)}{a^4}\varphi\varphi''
+2(m^2+\Delta m^2)
\frac{\varphi^2}{a^2}+\frac{\lambda+\Delta \lambda}{6a^4}
\varphi^4\nonumber\\&&
+4\Delta \tilde{\Lambda}+
\Delta \tilde{\alpha} H_{\mu}^{\mu}\, + \Delta \tilde{Z} G_\mu^\mu\, 
+T_{\mu}^{\mu}\,^{\rm q,fin}
 \pkt
\eea
The finite fluctuation parts of the trace of the energy-momentum tensor are
\bea \label{tmumufin}
T_{\mu}^{\mu}\,^{\rm q,fin}&=&(1-6\xi)
\frac{{\cal F}''_{\rm fin}}{2a^4}
+\frac{H}{a^3}(1-6\xi){\cal F}'_{\rm fin}
\nonumber\\
&&+\left[m^2+\frac{\lambda}{2a^2}\varphi^2-(1-6\xi)
\left(H^2-\frac {R}{6}\right)\right]\frac{{\cal F}_{\rm fin}}{a^2}
+T_{\mu}^{\mu}\,^{\rm add}\nonumber\\&&
\eea
where
\bea
T_{\mu}^{\mu}\,^{\rm add}&=
&T_{\mu}^{\mu}\,^{\rm ano}-\frac{1}{16\pi^2}\left(
\frac{V''}{6}+\frac{H^2M^2}{3a^2}+\frac{M^4}{2a^4}-\frac{V'}{3a^3}\right)
\\
&&
-\frac{1}{16\pi^2}\left(\xi-\frac 1 6\right)\left(
12\frac{H}{a^3}V'+\left(R+18H^2\right)M^2-\frac{1}{18} R^2-
\frac{\,^{(1)}H_{\mu\mu} }{36}\right)\nonumber
\kma
\eea
or, explicitly,
\bea
T_{\mu}^{\mu}\,^{\rm add}&=
&T_{\mu}^{\mu}\,^{\rm ano}
-\frac{3}{4\pi^2}\left(\xi-\frac 1 6 \right)^2\Biggl[H\frac{R'}{a}
+\frac 1 2 RH^2+\frac 1 8 R^2
\biggr]\nonumber\\
&&+\frac{1}{4\pi^2}\left(\xi-\frac 1 6\right)\Biggl[
-\frac{R''}{12a^2}-\frac{1}{6a}HR'-\frac{1}{4a^2}\lambda\varphi^2R
-\frac{1}{2}m^2R
\nonumber\\
&&-3\lambda\frac{H}{a^3} 
\varphi'\varphi+\frac{9}{4a^2}H^2\lambda\varphi^2-\frac{3}{2}H^2m^2
\Biggr]
-\frac{1}{32\pi^2}\left(m^2+\frac{\lambda}{2a^2}\varphi\right)^2\nonumber\\
&&-\frac{\lambda}{96\pi^2a^4}\left(\varphi'-aH\varphi\right)^2
-\frac{\lambda}{96\pi^2a^4}\varphi\varphi''-\frac{1}{288\pi^2}m^2R
\pkt
\eea
Comparing to Eq. (3.17) of \cite{Lindig:1998} 
some  differences in the 
coefficients can be absorbed into a different choice
of the finite parts of the renormalization constants.
The terms proportional to $\ln a(\tau)$ given there are included here
into $T_\mu^{\mu \,{\rm ren}}$ via the coefficients 
$\Delta \tilde \Lambda, \Delta \tilde Z$, and
$\Delta \tilde \alpha$; they are identical. 
We have checked, using MAPLE, that the energy 
momentum tensor is covariantly conserved. This is
a valuable cross check, as it relates the various terms in
$T_{tt}$ and $T_{\mu\nu}$ in a rather complex way. 


\section{Removing initial singularities}
\label{initial}


\setcounter{equation}{0}
The set of initial conditions used in the previous section leads,
after renormalization, to
singularities at $\tau=0$ in the remaining fluctuation integrals
occuring in the equation of motion for the inflaton and
in the energy-momentum tensor. This means that these
initial singularities affect the
Friedmann equations as well.
As already shown in \cite{Baacke:1998a} these singularities
can be removed by a suitable Bogoliubov transformation of the naive
initial state, i.e.,  the vacuum state of a Fock space based on the 
mode functions with the initial conditions Eq. (\ref{incons}).

Choosing such an initial state one finds, in the fluctuation integral
$\calf_{\rm fin}$ the terms
(\ref{flucintfin}):
\be
\calf_{\rm fin}^{\rm sing} =
 \intkd\Biggr[\frac{V'(0)}{8\Omk^4}\sin(2\Omk \tau)
-\frac{V''(0)}{16\Omk^5}\cos(2\Omk\tau)\Biggl]
\pkt\ee
While these are only nonanalytic as $\tau \to 0$, their
first and second derivatives are singular.
They occur in the energy-momentum tensor
via (\ref{tttfin}) and (\ref{tmumufin}).
The singular behaviour is given by     
\bea \nonumber
\calf_{\rm fin}^{\rm sing '}
&=&\intkd\Biggr[\frac{V'(0)}{4\Omk^3}\cos(2\Omk \tau)
+\frac{V''(0)}{16\Omk^4}\sin(2\Omk \tau)\Biggl]\\
&\stackrel{t\to 0}{\simeq}&
-\frac{1}{8\pi^2}\ln(2m\tau)V'(0)\kma\\
\calf_{\rm fin}^{\rm sing ''}&=&
\intk\Biggr[-\frac{V'(0)}{2\Omk^2}\sin(2\Omk \tau)
+\frac{V''(0)}{8\Omk^3}\cos(2\Omk \tau)\Biggl]\nonumber\\
&\stackrel{t\to 0}{\simeq}&-\frac{1}{16\pi^2 \tau}V'(0)+\frac{1}{16\pi^2}
\ln(2m\tau)V''(0)\pkt
\eea
A Bogoliubov-transformed initial state (which is the Heisenberg
state of the system)  can be defined by
requiring
\be
\left[a(\bfk)-\rho_ka^\dagger(\bfk)\right]|i\rangle =0 \pkt
\ee
If the fluctuation integral, the energy and the pressure are
computed by taking the trace with respect to this
state the functions $U_k(\tau)$ get replaced by
\be
F_k(\tau)=\cosh(\gamma_k)U_k(\tau)+e^{i\delta_k}
\sinh(\gamma_k)U_k^*(\tau) \kma
\ee
where $\gamma_k$ and $\delta_k$ are defined by
the relation
\be
\rho_k=e^{i\delta}\tanh(\gamma_k)\pkt
\ee
The fluctuation integral then becomes
\bea 
\calf(\tau)&=&\intk |F_k(\tau)|^2 \\ \nonumber
 &=&
\intk \left\{\cosh (2 \gamma_k(\tau)) |U_k(\tau)|^2+
\sinh (2\gamma_k)\re\left( e^{-i\delta_k}U^2_k(\tau)\right)\right\}\pkt
\eea
Expanding as before we find
\bea \nonumber
\calf (\tau)&=&\intk\left\{
\cosh(2\gamma_k)
\left[1-\frac{V(\tau)}{2\Omega_0^2}+\frac{V(0)}{2\Omega_0^2}\cos(2\Omega_0 \tau)
+\frac{V'(0)}{4\Omega_0^3}\sin(2\Omega_0 \tau)\right.\right.
\\ \nonumber&&\left.+\frac{V''(\tau)}{8\Omega_0^4}
-\frac{V''(0)}{8\Omega_0^4}\cos(2\Omega_0 \tau)-
\frac{1}{8\Omega_0^4}\calC(V''',\tau)
+ 2\re h_k^{(\overline{2})}+|h_k|^2 \right]
\\ &&+\sinh (2\gamma_k)\cos(\delta_k)\cos(2\Omega_0 \tau)
\label{fluc2}
\\ \nonumber
&&-\sinh(2\gamma_k)\sin(\delta_k)\sin(2\Omega_0 \tau)
+\sinh(2\gamma_k)\re e^{-2i\Omega_0 \tau-i\delta}
\left(2h_k+h_k^2\right)  \Bigg\}\pkt
\eea
Requiring that the terms proportional to
$V'(0)$ and $V''(0)$ vanish leads to the 
conditions 
\bea \label{deltadef}
\tan(\delta_k)&=&2\Omega_{k0}\frac{V'(0)}{V''(0)}\kma
\\ \label{gammadef}
\tanh (2\gamma_k)&=&
\frac{V''(0)}{8\Omega_{k0}^4}\left[1+\tan^2(\delta_k)\right]^{1/2} \pkt
\eea
For large $k$ this behaves as
\be
\gamma_k\stackrel{k\to\infty}{\simeq}\frac{|V'(0)|}{8\Omega_{k0}^3}
\pkt\ee
The factor $\cosh(2\gamma_k)$ is equal to $1$ for $\gamma_k=0$;
the difference w.r.t. $1$ is given by
\be \label{cosest}
\cosh(2\gamma_k)-1=2\sinh^2 (\gamma_k)\stackrel{k\to\infty}{\simeq}
\frac{V'^2(0)}{32\Omega_{k0}^4}\pkt
\ee
The  new terms proportional to $\sinh(2\gamma_k)$ behave as behave 
as
\be \label{sinest}
\sinh(2\gamma_k)\stackrel{k\to\infty}{\simeq}
 \frac{|V'(0)|}{4\Omega_{k0}^3}\pkt
\ee
The dimensionally regularized fluctuation integral (\ref{fluc2})
takes, after cancellation of the singular integrals induced
by Eqs. (\ref{deltadef}) and (\ref{gammadef}),
the form
\bea 
\calf_{\rm reg} (\tau)
&=&-\frac{m_0^2}{16\pi^2}(L_0+1)
-\frac{V(\tau)}{16\pi^2}L_0
\nonumber \\ 
&&+\intk\left\{
\sinh^2(\gamma_k)\left[1-\frac{V(\tau)}{2\Omk^2}\right]\right.
\nonumber \\
&&+\cosh(2\gamma_k)\left[\frac{V''(\tau)}{8\Omk^4}
\frac{1}{8\Omk^4}\calC(V''',\tau)
+ 2\re h_k^{(\overline{2})}+|h_k|^2 \right]
 \nonumber \\
&&+\sinh(2\gamma_k)\re e^{-2i\Omk \tau-i\delta}
\left(2h_k+h_k^2\right)  \Bigg\}\pkt
\eea
Using (\ref{cosest}) and (\ref{sinest}) we see that the Bogoliubov 
transform does not affect the ultraviolet divergences, and that,
therefore, the renormalization procedure remains unchanged. 

The structure of the equation of motion and the energy-momentum 
tensor remains the same after the
Bogoliubov transformation, if the fluctuation integral
$\calf_{\rm fin}$ is replaced by
\bea
\calft_{\rm fin}&=&-\frac{m_0^2}{16\pi^2}+\intkd\frac{1}{2\Omega_0}\Biggl\{
 2 \sinh^2(\gamma_k)\left[
1-\frac{V(\tau)}{2\Omega_{k0}^2}
\right]\nonumber\\
&&+\cosh(2\gamma_k)K_1(k,\tau)+\sinh(2\gamma_k)K_2(k,\tau)
\Biggr\}
\kma \eea
where we have introduced the functions:
\bea
K_1(k,\tau)&=&
\left[\frac{V''(\tau)}{8\Omega_{k0}^4}-\frac{{\cal C}(V''',\tau)}{8\Omega_{k0}^4}
+2\re h^{(\overline{2})}_k+|h_k|^2\right]\kma\\
K_2(k,\tau)&=&\re\left\{
e^{-2i\Omega_{k0}\tau-i\delta}\left(2h_k+h_k^2\right)
\right\}
\pkt \eea
Similarly, ${\cal E}_{\rm kin,fin}$ becomes
\bea
\tilde{{\cal E}}_{\rm kin,fin}&=&\frac{1}{a^4}\intkd\frac{1}{2\Omega_0}\Biggl\{
2\sinh^2(\gamma_k)\left[2\Omega_0^2
+\frac{V^2(\tau)}{4\Omega_0^2}\right]
\\ \nonumber
&&+\sinh (2\gamma)K_3(k,\tau)+K_4(k,\tau)\cosh^2(2\gamma_k)
\Biggr\}-\frac{3M_0^4}{128\pi^2a^4}
\kma \eea
where $K_3(k,\tau)$  denotes
\bea
K_3(k,\tau)&=&\re\left\{
e^{-2i\Omega_{k0}\tau-i\delta}\left[{h'}_k^2-2i\Omega_{k0}\left(1+h_k\right){h'}_k\right]
\right\}\kma\\
K_4(k,\tau)&=&\frac{1}{2}\Biggl[|h_k^{(\overline{1})'}|^2
-\frac{V^2(\tau)}{4\Omega_0^2}\Biggr]
\pkt \eea
With these replacements in the equation of motion and in the
energy-mo\-men\-tum tensor the inflaton eqution of motion and the Friedmann
equations are ultraviolet finite {\em and} free of initial singularities.
We should mention that this initial state is not determined
uniquely. Any state or density matrix based on this new
`vacuum state' is admissible as long as the spectrum of coherent or
incoherent exictations decreases stronger than the one parametrized by
the Bogoliubov angles $\gamma_k$ and $\delta_k$.  

We should like to mention that fixing the initial conditions 
results in a selfconsistency problem, as the
finite, modified fluctuation
integrals in the equation of motion and in the
energy-momentum tensor do not vanish at $\tau=0$. They depend on the
other initial parameters $H(0),R(0),\varphi(0),\varphi'(0)$,
and $\varphi''(0)$, and vice versa.
For the typical parameter sets, this selfconsistency problem
can be solved by an iteration which converges quickly.


\section{Large $N$ model}
\setcounter{equation}{0}
\subsection{General formalism}
We now consider the $O(N)$ $\sigma$ model defined by the 
Lagrangian 
\be
\label{lagrange}
{\cal L}=\frac{1}{2a^4}\partial_{\mu} \phi^i\partial^{\mu}\phi^i
-\frac{1}{2a^2}m^2\phi^i\phi^i
-\frac \xi 2  R \phi^i\phi^i-\frac{\lambda}{4Na^4}(\phi^i\phi^i)^2
\pkt \ee
whith $N$ real scalar fields $\phi^i, i=1,..,N$.
The nonequilibrium state of the system is characterized
by a classical expectation value which we take in 
the direction of $\phi_N$. We split the field
into its expectation value, or mean field, $\phi$ and 
the quantum fluctuations
$\psi$ via
\begin{equation}
\label{erw}
\phi^i(\bfx,\tau)=\delta^i_N \sqrt{N} \phi(\tau)+\psi^i(\bfx,\tau)\pkt
\end{equation}
In the large-$N$ limit one neglects, in the Lagrangian,
all terms which are not of order $N$. In particular
terms containing the fluctuation $\psi^N$ of
the component $\phi^N$ are at most of order $\sqrt{N}$ and are
dropped, therefore. This is in contrast to the Hartree approximation
where the fluctuations of $\phi_N$ are included. The fluctuations of
the other components are identical, their summation produces
factors $N-1=N(1+O(1/N))$. 
The quantum fluctuations are again decomposed into mode functions
via (\ref{fieldex}). 
The nonequilibrium equations of motion for the field $\phi(\tau)$ 
and of the mode functions $U_k(\tau)$ have been derived by
various authors \cite{Boya:1994,Cooper:1994}.
 These equations 
differ from those of the
one-loop approximation by the fact that the fluctuation integral
not only modifies the mass of the mean field, but also 
the one of the quantum fluctuations.   

The equation of motion for the mean field can be written as
\be \label{lnmean}
\phi''(\tau)+M^2(\tau)\phi(\tau)=0
\kma\ee
the one for the mode functions as
\be
U''_k(\tau)+\left[k^2+M^2(\tau)\right]U_k(\tau)=0
\kma
\ee \label{lnmodes}
with the same effective mass
\bea
M^2(\tau)&=&a(\tau)^2\left[m^2+\delta m^2+\left(\xi+\delta\xi-\xi_n
\right)R(\tau)\right]\nonumber\\
&&+ (\lambda+\delta
 \lambda)a(\tau)^{
\epsilon}\left(\phi(\tau)^2+\calf(\tau)\right)
\pkt\eea
We have included here the renormalization counter terms.
$\calf(\tau)$ is again the fluctuation integral
\be
\calf(\tau)=\intk \frac{1}{2\Omega_0}|U_k|^2
\pkt
\ee 
We restrict ourselves to a temperature $T=0$ system, here.
Generalization to thermal systems is straightforward
\cite{Baacke:1998b}.
As in the one-loop case we rewrite the mode equation in the form
\begin{equation} \label{lnudgl}
\left[ \frac{d^2}{d\tau^2}+
\Omega_{k0}^2\right]U_k(\tau)=-V(\tau)U_k(\tau)\; ,
\end{equation}
introducing, thereby,  the time dependent potential
\begin{equation} \label{lnV}
V(\tau)=M^2(\tau)-M^2(0)\; ,
\end{equation}
 The ``initial mass'' $M(0)=m_0$ will be determined below, as
a solution of a gap equation.
We define the time-dependent frequency $ \Omkt$
via
\be
\Omega_k^2(\tau) = k^2 +M^2(\tau)
\pkt\ee
In the large-$N$ limit, and in $n$ dimensions  the energy density 
is given by
\bea
T_{tt}&&=\frac{1}{a^{2-\epsilon}}\Biggl\{
\frac{1}{2a^2}\varphi'^2+\frac 1 2 (m^2+\delta m^2)
\varphi^2+\frac{(\lambda+\delta\lambda)(a\mu)^{
\epsilon}}{4a^2}\varphi^4\nonumber\\
&&+2(n-1)\left(\xi+\delta\xi-\xi_n\right)
\left(\frac{H}{a}\varphi\varphi'-\frac 1 4 
(n-2)H^2\varphi^2\right)-\frac{\lambda+\delta\lambda}{4a^{2-\epsilon}}\calf^2
\Biggr\}\nonumber\\
&&+\delta \tilde{\Lambda}+\delta\tilde{Z}G_{tt}\nonumber
+\delta \tilde{\alpha} 
\,^{(1)}H_{tt}
+\delta \tilde{\beta} 
\,^{(2)}H_{tt}
+\delta \tilde{\gamma} H_{tt}
\\
&&+\frac{1}{a^{2-\epsilon}}\intk\Biggl\{\frac{|U_k'|^2}{2a^2}+\frac{1}{2a^2}
\Omega_k(\tau)^2|U_k|^2\nonumber\\
&&+(n-1)\left(\xi+\delta\xi-\xi_n\right)
\left[\frac{H}{a}\frac{d}{d\tau}|U_k|^2-
\frac 1 2\left((n-2)H^2+\frac{R}{(n-1)}
\right)
|U_k|^2\right]\Biggr\}\kma\nonumber\\
\eea
and the trace of the energy-momentum tensor takes the form
\bea
T^{\mu}_{\mu}&=&\frac{1}{ a^{2-\epsilon}}\Biggl\{
2(n-1)\left(\xi+\delta\xi
-\xi_n\right)\left[
\frac{\varphi'}{a}-\frac 1 2 (n-2)H\varphi\right]^2\nonumber\\
&&+2(n-1)(\xi+\delta\xi)
\frac{\varphi\varphi''}{a^2}+n\left[\frac 1 2 (m^2+\delta m^2)\varphi^2+
\frac{(\lambda+\delta\lambda)(a\mu)^{\epsilon}}{24a^2}\varphi^4\right]
\Biggr\}\nonumber\\&&+n\delta \tilde{\Lambda}+ \delta\tilde{Z} G_\mu^\mu
+\delta \tilde{\alpha} 
\,^{(1)}H_{\mu}^\mu
+\delta \tilde{\beta} 
\,^{(2)}H_{\mu}^\mu
+\delta \tilde{\gamma} H_{\mu}^\mu
\nonumber\\
&&+\frac{1}{a^{2-\epsilon}}\intk\Biggl\{
\frac 1 2 \left[n-2-2(n-1)(\xi+\delta\xi)\right]\Biggl[
-\frac{|U_k'|^2}{a^2}+\frac{\Omega_k(\tau)}{a^2}|U_k|^2\nonumber\\
&&+\frac 1 2 (n-2)\frac H a \frac{d}{d\tau}|U_k|^2-\frac 1 4 (n-2)^2\left(
H^2-\frac{R}{(n-1)(n-2)}\right)|U_k|^2
\Biggr]\nonumber\\
&&+\left(m^2+\delta m^2+\frac{(\lambda+\delta\lambda)
(a\mu)^{\epsilon}}{2a^2}\right)|U_k|^2
\Biggr\}
-\frac{1}{4a^{4-2\epsilon}} (n-4)(\lambda+\delta\lambda)\calf^2\pkt
\nonumber\\
\eea


\subsection{Renormalization}

The way in which the renormalization counter terms
are determined in the large-$N$ case
has been described in detail in \cite{Baacke:1998b}. 
We will closley follow this approach.
As in the one-loop case
use  the expansion of the mode functions in order to single
out the divergent parts:
\be\label{flucintdiv}
\calf=-M^2(\tau)x_0-\frac{m_0^2}{16\pi^2}+\calf_{\rm fin}
\ee
with
\be
x_0=\frac{1}{16\pi^2}\left\{\frac{2}{\epsilon}
+\ln{\frac{4\pi\mu^2}{M^2(0)}}-\gamma\right\}
\; .
\ee
The finite part of the fluctuation integral reads
\begin{equation}\label{lnflucintfin}
{\cal F}_{\rm fin}=\intk\frac{1}{2\Omega_{k0}}
\left(
\frac{1}{2\Omega_{k0}}{\cal C}(V',\tau)
+2{\rm Re }\,{h}_k^{(\overline{2})}+|h_k^{(\overline{1})}|^2\right)\; ,
\end{equation}
The renormalization conditions for mass, coupling to the Ricci scalar
and coupling constant 
are obtained from the requirement that the frequencies which  
appear in the mode equations are finite, 
i.e. that $M^2(\tau)$ is a finite quantity:
\bea
a^2\left[m^2+\delta m^2+(\xi+\delta\xi-\frac 1 6)R\right]+ (\lambda+\delta
 \lambda)\left({\phi^2}+{\calf}\right)\nonumber
&&\\=
a^2\left[m^2+(\xi-\frac 1 6)R\right]+\lambda \left(\phi^2
-\frac{m^2_0}{16\pi^2}+\calf_{\rm fin}\right)
\eea
From this condition, and using Eq. (\ref{flucintdiv}),
 we find the following counter terms
\bea
\delta \lambda&=&\frac{\lambda^2 x}{1-\lambda x}\kma\\
\delta m^2&=&\frac{m^2}{1-\lambda x}\left(x+\rho \lambda\right)\kma\\
\delta \xi&=&\frac{\lambda}{1-\lambda x}\left[\left(\xi-\frac 1 6\right)x
+\sigma\right]
\pkt \eea
The occurence of the free parameters $\rho,\sigma$ shows that
the counter terms are not determined uniquely; the choice of
these parameters corresponds to the freedom of an independent choice
of renormalization conventions. 
 
We obtain for $M^2$ the manifestly finite expression:
\bea \label{mtau}
M^2(\tau)&=&\caln(\tau)\Biggl\{a(\tau)^2\left[m^2+\lambda\rho
+(\xi-\frac 1 6+\lambda\sigma)R(\tau)\right]\nonumber\\
&&+\lambda\left[\phi(\tau)^2
-\frac{m^2_0}{16\pi^2}-\frac{a(\tau)^2R(\tau)}{288\pi^2}+\calf_{\rm fin}(\tau)\right]\Biggr\}
\pkt\eea
Here
\be
\caln(\tau)=
\frac{1}{1-\lambda(x-a(\tau)^{\epsilon}x_0)}
=\frac{1}{1+\frac{\lambda}{16\pi^2}\ln{\frac{m^2a(\tau)^2}{m^2_0}}}
\kma \ee
and
\be
x=\frac{1}{16\pi^2}\left\{
\frac{2}{\epsilon}+\ln{\frac{4\pi\mu^2}{m^2}}-\gamma\right\}\; .
\ee
We are now able to determine $m_0^2$. Taking Eq. (\ref{mtau}) at
$\tau=0$ we obtain an implicit equation for $m_0=M(0)$, the gap
equation
\bea\label{mzero}
m_0^2&=&\caln\Biggl\{a^2(0)\left[m^2+\lambda\rho
+(\xi-\frac 1 6+\lambda\sigma)R(0)\right]
\\ \nonumber
&&+\lambda\left[\phi^2
-\frac{m^2_0}{16\pi^2}-\frac{a^2(0)R(0)}{288\pi^2}\right]\Biggr\}
\pkt\eea
With Eqs. (\ref{mtau}) and (\ref{mzero}) the
 time dependent potential
$V(\tau)=M^2(\tau)-m_0^2$ can be expressed in terms of finite quantities.

Renormalization of the energy-momentum tensor proceeds analogously. 
We insert the perturbative
expansion of the mode functions and determine the 
remaining counter terms so as to render the tensor
finite. We require them again to be
independent of time and of the initial conditions.
The energy density  then is given by
\bea
{\cal E}&=&\frac{1}{a^{2-\epsilon}}\left\{\frac{1}{2a^2}\varphi'^2
+\frac{1}{2} (m^2+\delta m^2)\varphi^2
+\frac{\lambda+\delta \lambda}{4a^{2-\epsilon}}\varphi^4\right\}
\nonumber \\&&
+\delta \tilde{\Lambda}+\delta \tilde{\alpha}\,^{(1)} H_{tt}
\delta \tilde{\beta}\,^{(2)} H_{tt}+\delta \tilde{\gamma} H_{tt}
+\delta\tilde{Z} G_{tt}\nonumber\\
&&+2(n-1)(\xi+\delta\xi-\xi_n)\left(-\frac1 4(n-2)
\frac{H^2}{a^{2-\epsilon}}\varphi^2+\frac{H}{a^{3-\epsilon}} \varphi\varphi'
\right)+{\cal E}_{\rm kin,fin}
\nonumber\\
&&+\frac{1}{a^{2-\epsilon}}\Biggl\{
-\frac{M^2(\tau)^2x_0}{4a^2}-\frac{m_0^2V(\tau)}{32\pi^2a^2}
+\frac{V(\tau)\calf_{\rm fin}}{2a^2}+
\nonumber\\
&&-\frac{1}{2}(n-1)
\left(\xi+\delta\xi-\xi_n\right)\left(\frac{R}{2(n-1)}
 + \frac 1 2 (n-2)H^2\right)\times\nonumber\\
&&\hspace{6cm}\times\left(-M^2(\tau)x_0-\frac{m_0^2}{16\pi^2}+\calf_{\rm fin}
\right)\nonumber\\
&&
+\frac{H}{a}\left(n-1\right)(\xi+\delta\xi-\xi_n)
\left(-V'(\tau)x_0+\calf_{\rm fin}'
\right)\nonumber\\
&&-\frac{\lambda+\delta\lambda}{4a^{2-\epsilon}}
\left(-M^2(\tau)x_0-\frac{m_0^2}{16\pi^2}+\calf_{\rm fin}\right)^2\Biggr\}
\kma \eea
explicitly displaying the divergent parts.
We have defined
\be
{\cal E}_{\rm kin,fin}=-\frac{3m_0^4}{128\pi^2a^4}+
\frac{1}{2 a^4} \intkd\frac{1}{2\Omk} 
\Biggl[|h_k^{(\overline{1})'}|^2
-\frac{V^2(\tau)}{4\Omk^2}\Biggr]
\pkt \ee
After inserting the expressions 
for the coefficients $\delta \lambda, \delta m^2$ and $\delta
\xi$ we can now fix the coefficients of
the higher derivative counter terms,
the cosmological
 constant  and the wave function renormalization so as to
obtain a finite expression for the energy density. 
This is a very tedious exercise; using MAPLE we
find 
\bea
\delta \alpha&=&-\frac{\left(\xi-
\frac 1 6+\lambda\sigma \right)^2x}{2(1-\lambda x)}\kma \\
\delta \Lambda&=&\frac{xm^4\left(1+\frac{\lambda\rho}{m^2}\right)^2}
{4(1-\lambda x)}\kma \\
\delta Z&=&\frac{xm^2\left(1+\frac{\lambda\rho}{m^2}\right)\left(
\xi-\frac 1 6 +\lambda\sigma
\right)}{1-\lambda x}\kma
\eea
and renormalized energy density 
\begin{eqnarray} \label{lntttren}
{\cal E}_{ren}&=&
\frac{\varphi'^2}{2a^4}+\caln(m^2+\lambda\rho)\frac{\varphi^2}{2a^2}
+\frac{\caln \lambda}{4a^4}\varphi^4\nonumber
\\&&-\caln\left(6\xi-1+6\lambda\sigma\right)\left(
\frac{H^2}{2a^2}\varphi^2-\frac{H}{a^3}
\varphi\varphi'
\right)+\Delta \tilde{\Lambda}+\Delta \tilde{\alpha} H_{tt}
+\Delta\tilde{Z} G_{tt}
\nonumber\\
&&+{\cal E}_{\rm kin,fin}
+\frac{1}{2a^4}\left[V(\tau)+\caln\lambda
\frac{a^2 R}{288\pi^2}\right]
\left(\tilde{ \cal F}_{\rm fin}-\frac{m_0^2}{16\pi^2}
\right)\nonumber\\
&&-\frac{\caln}{2a^2}(6\xi-1+6\lambda\sigma)\left(\frac R 6+H^2\right)
\left(\tilde{\cal F}_{\rm fin}-\frac{m_0^2}{16\pi^2}\right)
\nonumber\\
&&+ \caln(6\xi-1+6\lambda\sigma)\frac{ H}{2a^3} 
\tilde{\cal F}'_{\rm fin}
-\frac{\caln\lambda}{4a^4}\left(\calft_{\rm fin} 
-\frac{m_0^2}{16\pi^2}\right)^2\nonumber\\
&&+T_{tt}^{\rm ano}
+\frac{H}{96\pi^2a^3}\left(aHM^2-V'\right)-\frac{\caln}{331776\pi^4}
\lambda R^2\nonumber\\
&&+\frac{\caln}{16\pi^2}\left(\xi-\frac 1 6\right)\left(
\frac{\,^{(1)}H_{tt}}{36}+\frac{R^2}{72}+\frac{6H^2M^2}{a^2}\right)
\pkt
\end{eqnarray}
The coefficients of the finite parts of the counter terms are:
\bea
\Delta\tilde{\alpha}&=&\frac{\caln}{1152\pi^2}
\left(6\xi-1+6\lambda\sigma\right)^2\ln{\frac{m^2a(\tau)^2}{m_0^2}}\kma\\
\Delta\tilde{Z}&=&\frac{\caln}{432\pi^2}\left(m^2+\lambda\rho\right)\left(6\xi-1
+6\lambda\sigma\right)\ln{\frac{m^2a(\tau)^2}{m_0^2}}\kma\\
\Delta\tilde{\Lambda}&=&-\frac{\caln}{64\pi^2}\left(m^2+\lambda\rho\right)^2
\ln{\frac{m^2a(\tau)^2}{m_0^2}}
\pkt
\eea
The renormalized trace of the energy-momentum tensor does
not require any further counter terms, we find
\bea \label{lntmumuren}
T_{\mu}^\mu&=&
\caln\left(6\xi-1+6\lambda\sigma\right)\left(
\frac{\varphi'^2}{a^4}+  \frac{\varphi\varphi''}{a^4}
+\frac{H^2}{a^2}\varphi^2-2\frac{H}{a^3}\varphi\varphi'
\right)+\frac{\varphi\varphi''}{a^4}\nonumber\\
&&+2\caln(m^2+\lambda\rho)
\frac{\varphi^2}{a^2}+\frac{\caln\lambda}{a^4}
\varphi^4
+4\Delta \tilde{\Lambda}+
\Delta \tilde{\alpha} H_{\mu}^{\mu} + \Delta \tilde{Z} G_\mu^\mu 
\nonumber \\
&&
-\caln(1-6\xi-6\lambda\sigma)
\frac{\tilde{\cal F}''_{\rm fin}}{2a^4}
-\caln(6\xi-1+6\lambda\sigma)\frac{H}{a^3}\tilde{\cal F}'_{\rm fin}
\nonumber\\
&&+\caln
\left(m^2+\lambda\rho+\frac{\lambda}{a^2}\varphi^2\right)\left(
\frac{\tilde{\cal F}_{\rm fin}}{a^2}-\frac{m_0^2}{16\pi^2a^2}\right)
\nonumber\\
&&
+\caln(6\xi-1+6\lambda\sigma)
\left(H^2-\frac {R}{6}\right)\left(
\frac{\tilde{\cal F}_{\rm fin}}{a^2}-\frac{m_0^2}{16\pi^2a^2}\right)
\nonumber\\
&&
+T_{\mu}^{\mu}\,^{\rm ano}-\frac{1}{16\pi^2}\left(
\frac{V''}{6}+\frac{H^2M^2}{3a^2}+\frac{M^4}{2a^4}-\frac{V'}{3a^3}\right)
\nonumber \\
&&
-\frac{\caln}{16\pi^2}\left(\xi-\frac 1 6\right)\left(
12\frac{H}{a^3}V'+\left(R+18H^2\right)M^2-\frac{1}{18} R^2-
\frac{\,^{(1)}H_{\mu\mu} }{36}\right)\nonumber
\\
&&
-\frac{\caln}{18\pi^2}\frac{\lambda R}{a^2}\left(
\frac{m_0^2}{16\pi^2}-\tilde{\cal F}_{\rm fin}+\frac{a^2 R}{288\pi^2}
\right)\pkt
\end{eqnarray}


\subsection{Removing initial singularities}


We again have to remove the initial singularities.
The construction of the initial state, and the modified expressions
are analogous to those of section \ref{initial}. Of course
for the potential $V(\tau)$ one has to use Eqs.
(\ref{lnV}), (\ref{mtau}), and (\ref{mzero}).
As in the one-loop case the initial parameters have to be determined
in a selfconsistent way, here in addition one has to solve
the gap equation.


\section{Numerical results}


We have numerically implemented the formalism developed
in the previous sections. We present here some
results of the numerical computations.
We restrict the presentation to results for the
$O(N)$ $\sigma$ model.

We consider the coupled evolution of the scale parameter
via the renormalized 
Friedmann equations (\ref{Friedtt}), (\ref{Friedtr}) 
with the energy-momentum tensor given in Eqs.
(\ref{lntttren}) and (\ref{lntmumuren}),  of 
the inflation field via Eq. (\ref{lnmean}), and of the 
quantum modes using Eq. (\ref{lnmodes}). The quantum back reaction 
on the inflaton field and the scale parameter is included
from the initial time on, this start having made possible
by the modification of the initial state.

We show results for different sets of parameters,
choosing values that are considered realistic for chaotic
inflation, similar to those used in \cite{Ramsey:1997}. 
We do not consider the evolution during inflation,
but only during the reheating period. This means that the initial
amplitude of the inflaton field $\phi(0)$ is less or at most of the
order of the Planck mass. During  inflation the low momentum
modes increase exponentially, requiring 
a different numerical or semianalytic approach 
\cite{Boya:1998I,Boya:1997C2,Cormier:1998}.

We set  $\lambda=10^{-12}$ throughout. This is a value 
considered to be realistic. The inflaton mass is currently
estimated to be of the order $m \simeq 10^{-6}\cdot M_P$. 
With such a value fluctuations do not develop, even if
$\varphi(0)\simeq M_P$, and the evolution is
dominated by the classical amplitude. We therefore consider smaller 
values of the inflaton mass, we also vary the initial 
field amplitude.
In one group of parameter sets we fix the ratio
$\varphi(0)/m$ and vary the ratio $m/M_P$, as it was done in
\cite{Ramsey:1997}. In another set we start with
$\varphi(0)=2 M_P$ and vary the ratio $m/M_P$.
The parameters of the various sets are displayed in Table 1.

In the first group with the  parameters sets $1, 2$ and $3$ we fix
$\varphi(0)/m=2\cdot 10^7 $ and choose $m/M_P = 10^{-12}, 10^{-8},
$ and $10^{-7}$, respectively. This entails that the initial amplitudes vary
along with the masses.
The absolute time scale is determined by setting $m=1$ in the numerical 
computations. So  for all Figures 
which refer to this data set the abscissa is $m \tau$. This also determines
the units of the various physical quantities.

 For first parameter set the initial amplitude 
is extremely small, and so is the back reaction on the scale parameter.
We have essentially the situation of Minkowski space-time.
 The scale parameter stays almost constant.
The evolution of the field amplitude $\varphi$ is shown in 
Fig. 1.  
The fluctuations develop until $\calf_{\rm fin} \simeq \varphi^2$ and then
have a stationary amplitude, as well-known from studies of the
large $N$ model \cite{Boya:1996T,Baacke:1998b}. 
We display, in Fig. 2, the fluctuation energy and the total
energy, which show a small asymptotic decrease due to the expansion.
The asymptotic value of the ratio $p/{\cal E}$ is found to be
$1/3$, corresponding to a radiation dominated ensemble.

For the second set the evolution of the scale parameter is
plotted in Fig. 3; it is
sizeable as to be expected for an 
initial amplitude $\varphi(0)=0.2 M_P$. The classical field amplitude,
as displayed in Fig. 4 ,
starts decreasing after $m \tau \approx 5$;  the asymptotic decrease is
stronger than expected for the large-time behaviour
in Minkowski-space time ($\tau^{-.27}$), and 
so is due, in part,  to the expansion.
The fluctuations,
diplayed in Fig. 5 develop again until the asymptotic
amplitude is reached,  at the same time they are red-shifted.
The ratio $p/{\cal E}$, shown in Fig. 6 becomes $\simeq 1/3$ 
(radiation dominated) after the fluctuations set in,
and decreases to zero asymptotically. This has similarly been found
in \cite{Boya:1997C2} for a similar parameter set.
The fluctuation energy becomes of the same order, as the total energy,
see Fig. 7  . We finally plot, in Fig. 8 , a typical 
energy distribution of the
produced quanta,  with the familar resonance band at 
low energy.

For parameter set $3$ the field amplitude starts at $2 M_P$. The
scale parameter develops strongly (see below). The 
fluctuations hardly develop,
they are redshifted immediately. The evolution is essentially
driven by the classical field, with $\langle p/{\cal E}\rangle \approx 0$.
We show the classical amplitude in Fig. 9., its decrease is mainly due
to the expansion. 

These results are qualitatively analogous to those 
of \cite{Ramsey:1997}, where
$\lambda$ was chosen $10^{-14}$. The main difference is in the
time scale for the built-up of fluctuations. In
 \cite{Ramsey:1997} the mass $m$ is of the order $\sqrt{\lambda} \phi(0)$,
while here the latter quantity is ten times bigger, and dominates
in the estimates for the time scales.

We now consider the second group of parameter sets, for
which the initial amplitude is fixed, $\varphi(0)=2 M_P$.
For the sets $3$, $4$, and $5$ the inflaton mass then is
chosen as $m/M_P= 10^{-7}, 10^{-8}$, and $10^{-12}$, 
respectively. Numerically we have put $M_P=10^7$ for all
parameter sets. So for all Figures which refer to
this data set, the abscissa is $10^{\-7}M_P\tau$, and the 
units for all physical quantities are fixed correspondingly.

The data set $3$ has been discussed above, it displays a strong expansion
and fluctuations play essentially no r\^ole. The situation changes with
decreasing inflaton mass, which for data set $4$ is smaller 
by a factor of $10$. The universe then still expands strongly, but
at the same time we find sizeable fluctuations. Still the fluctuation
energy only amounts to one part in $10^5$ of the total field energy.
For the much smaller masses value $m/M_P=10^{-12}$ the expansion
rate only goes down by a factor of about $5$, the fluctuation
energy is of the same order as the total energy.   

We show, in Fig. 10, the evolution of the scale parameter
for the data sets $3$, $4$, and $5$, all of which show a strong increase.
The computation of the time evolution for data set $3$ was stopped
at $\tau\simeq 9$ as the oscillation period in conformal time 
decreases, requiring very small time steps. The evolution becomes 
uninteresting in the present context, as the fluctuations are negligible.
In Fig. 11 we display the fluctuation energy for these 
parameter sets. We also plot the total energy for set $5$, 
the total energy for the other sets is almost identical, on the
logarithmic scale. It is clearly seen that for set $3$ the fluctuations
hardly develop and are redshifted immediately. For set $4$ with a
somewhat smaller inflation mass the fluctuations evolve but remain
on the level of one part in $1000$. Finally, for very small
masses the fluctuation energy increases to the same order of magnitude,
roughly $90 \%$ of the total energy.  

The computations were performed on
PentiumII PC's ,  the CPU time for one of the parameter sets
as presented above, was $4-8$ hours. The typical time step,
on the scale displayed in the Figures, was $10^{-3}$,
with an adaptive step-size control. The
covariant energy conservation was fulfilled to better than 
$1$ part in $10^5$.

\section{Outlook}

We have presented here the renormalized equations of motion for
a self-coupled scalar field in a conformally flat FRW universe including
the quantum back reaction in one-loop and large-$N$
approximations. We have applied this formalism to the
post-inflationary preheating period.  
Our results are consistent with those of
other authors if we choose the same parameter sets.
We found that for sufficiently light inflaton masses and initial
values in the range $\varphi \simeq M_P$ one can have 
a substantial growth of the energy of the fluctuations
at the same time as a substantial cosmological expansion.
Such low values of the mass are not in the range of
commonly accepted inflaton masses; however, this case could be relevant
if the quantum fluctuations are those of other fields.

We have not considered the inflationary stage itself.  
In this case, for $\xi=0$, the low momentum modes grow exponentially
due to a term $- R/6$ in the effective mass, even in the
absence of spinodal decomposition. This exponential growth
and the extreme red-shifting pose special numerical problems,
 so that semi-analytical techniques become
necessary. 
The collective evolution of the low momentum modes
(`zero mode assembly') indeed leads to an essential simplification
\cite{Boya:1997C2,Cormier:1998}.

We have considered here the quantum fluctuations of the inflaton
field itself. It would be interesting to couple other fields,
as e.g. fermion or gauge fields to the inflaton. This would allow
for a more general choice of parameters and could lead to 
interesting phenomena as e.g. fermionic preheating
\cite{Baacke:1998c,Greene:1998} 
 
It is known since some time \cite{Mukhanov:1992},
that gravitational gauge invariance
requires the inclusion of the scalar metric perturbations
on the same level as the inclusion of the scalar
field fluctuations. The resonant growth of such metric
perturbations in conjunction with those of the scalar
field has been considered recently \cite{Bassett:1999}.
It would be interesting to extend
the formalism presented here so as to include metric
perturbations.


\section*{Acknowledgements}

The authors thank D. Boyanovsky, H. de Vega and K. Heitmann
for useful discussions, and the theory group
of the LPTHE at the Universit{\'e} Pierre et Marie Curie, Paris,
for their warm hospitality. J.B. thanks
 the group T8 at the Los Alamos National Laboratory for their
hospitality,  E. Mottola and P. R. Anderson for useful
 discussions, and the Deutsche Forschungsgemeinschaft
 for partial support under contract Ba-703/6-1. C.P. thanks
 the Graduiertenkolleg ``Erzeugung und Zerf\"alle von 
Elementarteilchen'' at the University of Dortmund
 for financial support.


\newpage
\begin{center}{\bf \large Table Captions}
\end{center}
{\bf Table 1:} Parameter sets. We display the parameters
of the parameter sets $1-5$. We only show the various ratios.
The numerical values of the 
dimensionful quantities $m, M_P$, and $\varphi(0)$ in the
computer code are fixed by the numerical value $\varphi(0)=2\cdot 10^7$.
\\
\begin{center}
{\bf \large Figure Captions}
\end{center}
\noindent
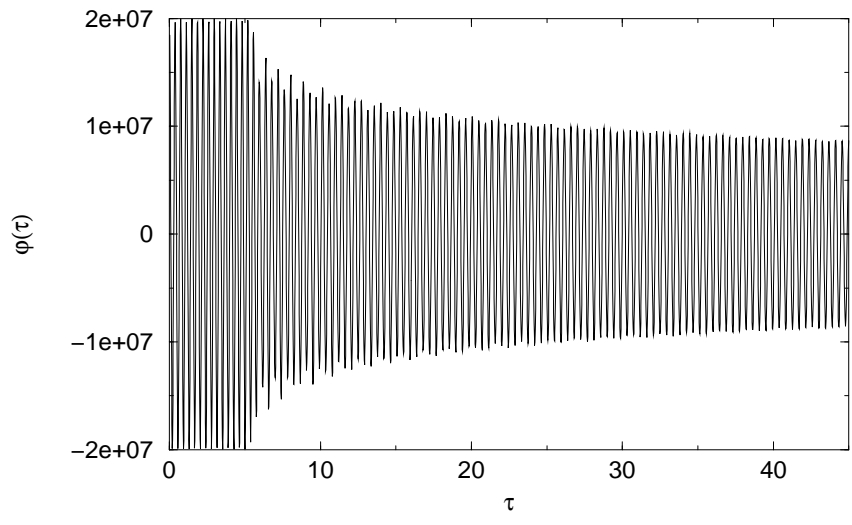
{\bf Fig. 1:} Classical amplitude 
$\varphi(\tau)$ for parameter set $1$. \\
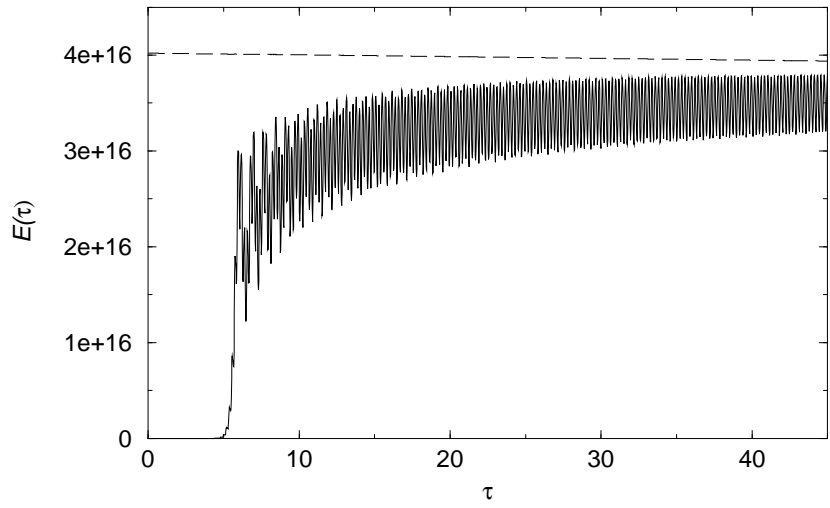
{\bf Fig. 2:} Fluctuation energy (solid line) and total energy 
(dashed line) for 
parameter set $1$. \\
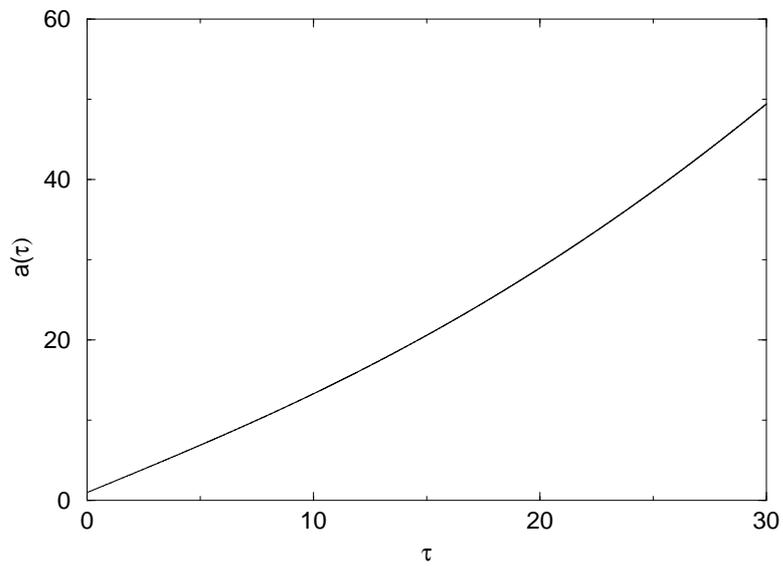
{\bf Fig. 3:} Evolution of the scale parameter $a(\tau)$ for
parameter set $2$. \\
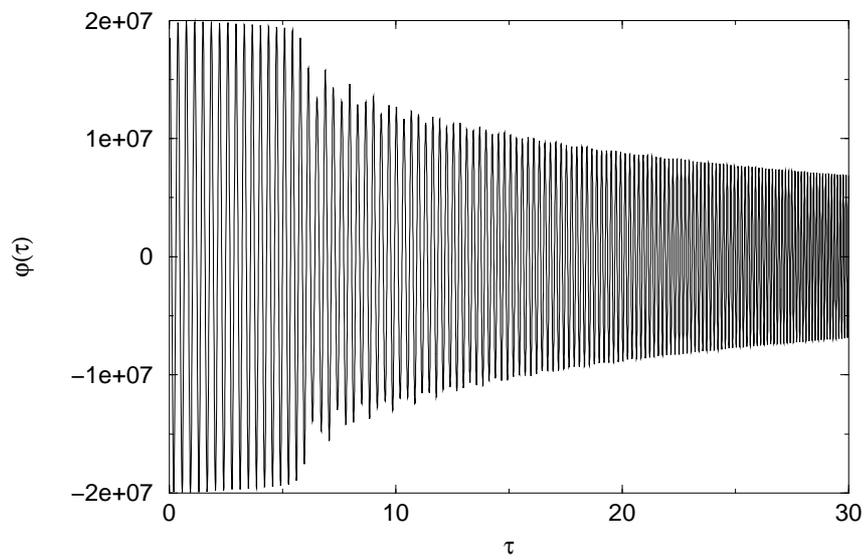
{\bf Fig. 4:} Classical amplitude $\varphi(\tau)$ for parameter set $2$. \\
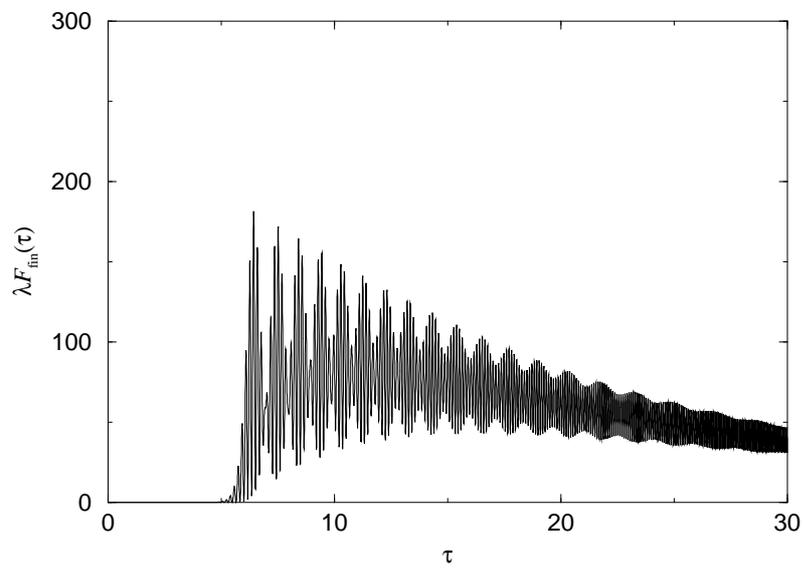
{\bf Fig. 5:} Fluctuation integral $\calf_{\rm fin}$ for parameter set $2$. \\
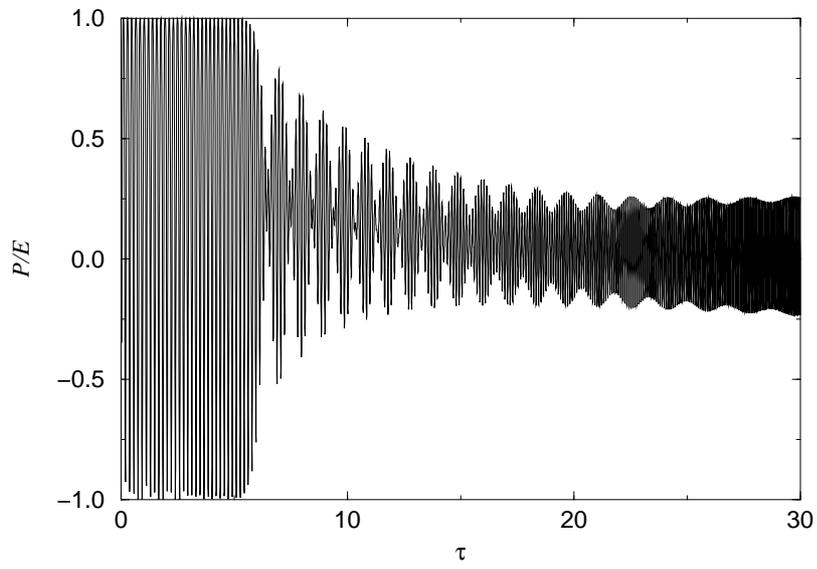
{\bf Fig. 6:} Ratio if pressure to energy density, $p/{\cal E}$, 
for parameter set $2$. \\
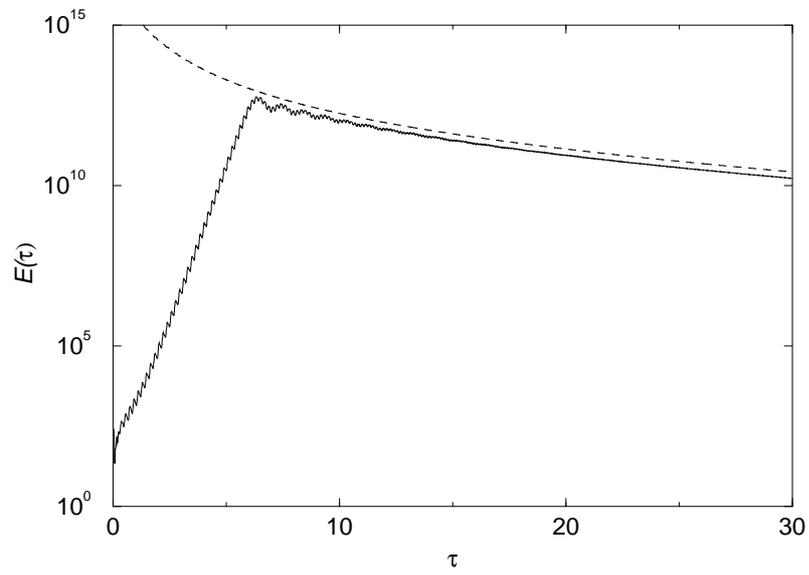
{\bf Fig. 7:} Fluctuation energy (solid line) and total energy (dashed line) for
parameter set $2$. \\
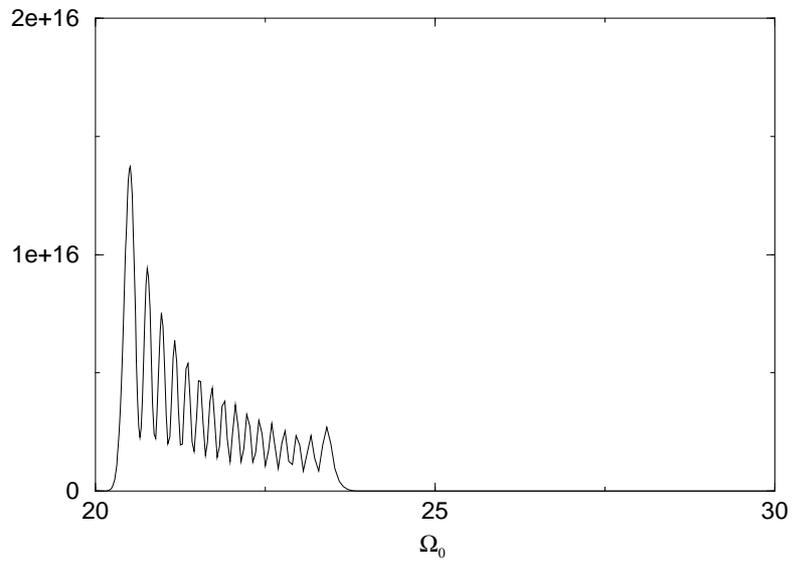
{\bf Fig. 8:} The energy distribution of the fluctuations at 
$\tau=20$ for parameter $2$. \\
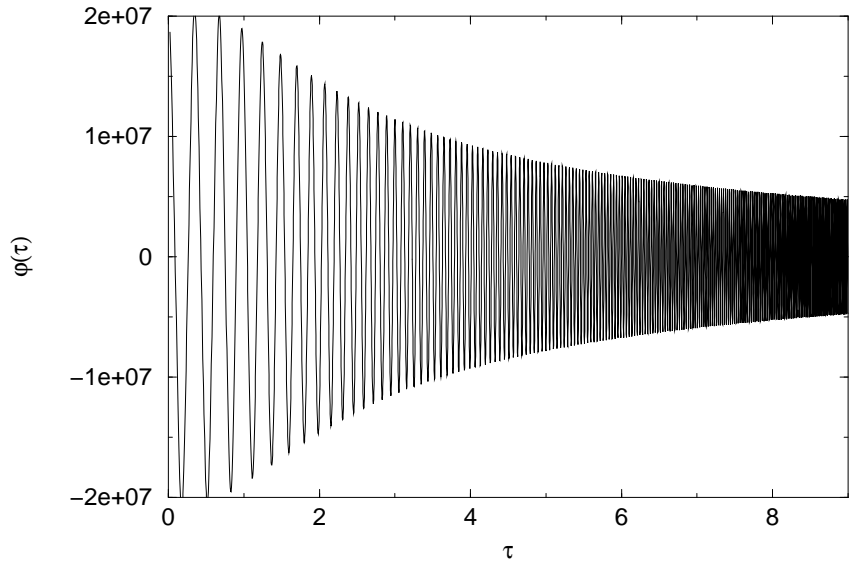
{\bf Fig. 9:} Classical amplitude $\varphi(\tau)$ for parameter
set $3$. \\
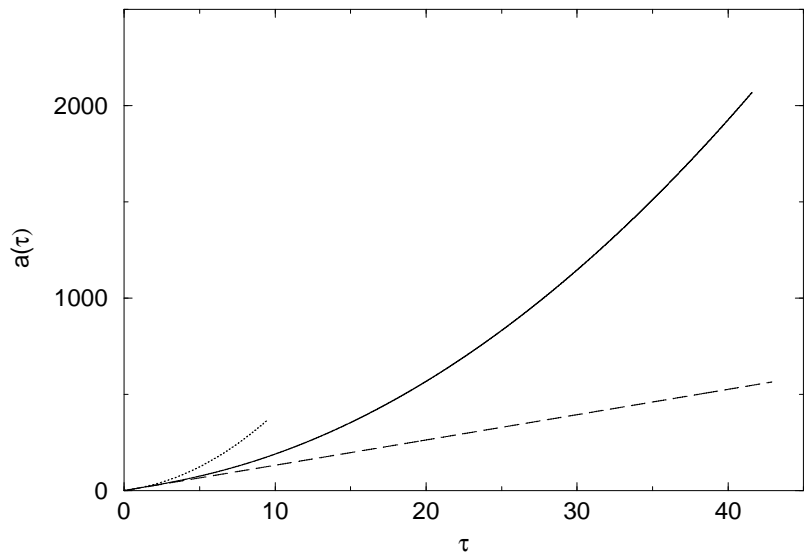
{\bf Fig. 10:} Evolution of the scale factor $a(\tau)$
for parameter set $3$ (dotted line),
set $4$ (solid line), and set $5$ (dashed line). \\
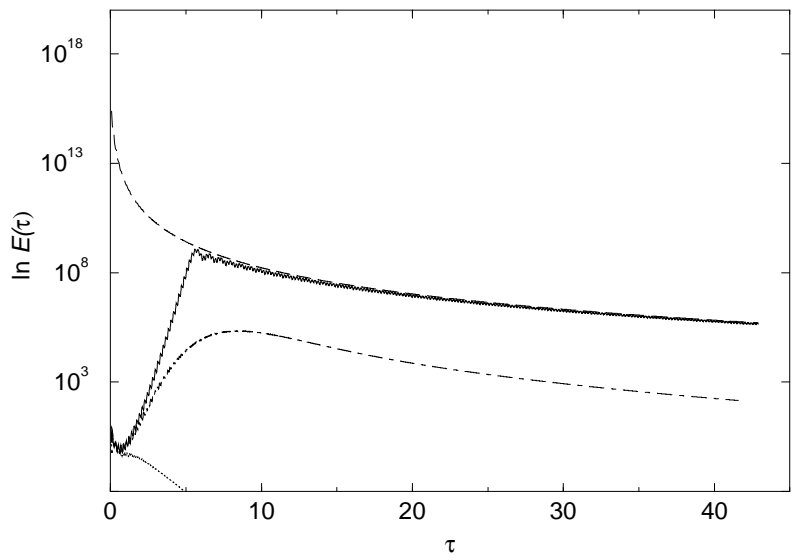
{\bf Fig. 11:} Evolution  of the fluctuation energy
for parameter set $3$ (dotted line), set $4$ (dash-dotted line),
and set $5$ (solid line). The total energy (dashed line) is the one
for parameter set $5$.
\newpage
\parbox{15cm}{
\begin{center}
\begin{tabular}{|c|c|c|c|}
\hline
set \# & $ \lambda$ & $m/M_P$ & $\varphi(0)/M_P$\\
\hline
$1$&$10^{-12}$&$10^{-12}$&$2\cdot10^{-5}$\\
$2$&$10^{-12}$&$10^{-8} $&$2\cdot10^{-1}$\\
$3$&$10^{-12}$&$10^{-7} $&$2$\\
$4$&$10^{-12}$&$10^{-8} $&$2$\\
$5$&$10^{-12}$&$10^{-12}$&$2$\\
\hline
\end{tabular} \end{center}}
\begin{center}
{\bf Table 1}
\end{center}

\end{document}